\documentclass[letterpaper]{mn2e}
\input{psfig}
\def\beq{\begin{equation}}
\def\eeq{\end{equation}}
\def\barray{\begin{eqnarray}}
\def\earray{\end{eqnarray}}

\def\gtsima{$\; \buildrel > \over \sim \;$}
\def\ltsima{$\; \buildrel < \over \sim \;$}
\def\prosima{$\; \buildrel \propto \over \sim \;$}
\def\gsim{\lower.7ex\hbox{\gtsima}}
\def\lsim{\lower.7ex\hbox{\ltsima}}
\def\simgt{\lower.7ex\hbox{\gtsima}}
\def\simlt{\lower.7ex\hbox{\ltsima}}
\def\simpr{\lower.7ex\hbox{\prosima}}

\newcommand{\apj}{ApJ}

\newcommand{\aj}{AJ}
\newcommand{\mnras}{MNRAS}

\newdimen\hssize
\newdimen\hdsize
\def\lesssim{\mathrel{\hbox{\rlap{\hbox{\lower4pt\hbox{$\sim$}}}\hbox{$<$}}}}
\def\gtrsim{\mathrel{\hbox{\rlap{\hbox{\lower4pt\hbox{$\sim$}}}\hbox{$>$}}}}

\newcommand{\fof}{{\scshape fof~}}
\newcommand{\subfind}{{\scshape subfind~}}

\begin{document}


\title[]{Cluster Galaxies Die Hard}
\author[S. M. Weinmann, G. Kauffmann, A. von der Linden, G. De Lucia]          
       {\parbox[t]{\textwidth}{
        Simone M. Weinmann$^{1}$\thanks{E-mail:simone@MPA-Garching.MPG.DE}, 
        Guinevere Kauffmann$^{1}$, Anja von der Linden$^{2}$,\\
        Gabriella De Lucia$^{3}$}\\
\vspace*{3pt}\\
$^1$Max-Planck Institut fuer Astrophysik, Karl-Schwarzschild-Str.1, Postfach 1317, 85741 Garching, Germany\\
$^2$Kavli Institute for Particle Astrophysics and Cosmology, Stanford 
University, 452 Lomita Mall, Stanford, CA 94305-4085, USA\\
$^3$Osservatorio Astronomico di Trieste INAF, Via Tiepolo 11, 34143
Trieste, Italy}


\date{}

\pubyear{2009}

\maketitle

\label{firstpage}


\begin{abstract}
We investigate how the specific star formation rates of galaxies
of different masses depend on  cluster-centric radius and
on the  central/satellite dichotomy in both field and cluster
environments. Recent data from a variety of sources, including
the cluster catalogue of von der Linden et al. are compared to the 
semi-analytic models of De Lucia \& Blaizot. We find that these models
predict too many passive  satellite galaxies in clusters, too few
passive central  galaxies with low stellar masses, and too many 
passive central galaxies  with high masses. We then outline a
series of modifications to the model  necessary to solve
these problems: a) Instead of instantaneous stripping of the external gas
reservoir after a galaxy becomes a satellite, the gas supply is assumed to 
decrease at the same rate that the surrounding halo 
loses mass due to tidal stripping, b) The AGN feedback efficiency is lowered
to bring  the fraction of massive passive centrals in better agreement
with the data.
We also allow for radio mode AGN feedback in satellite galaxies.
c) We assume that satellite galaxies residing in host
haloes
with masses below $10^{12} h^{-1}M_{\odot}$ do not undergo any stripping.
We highlight the fact that in low mass galaxies, the external
reservoir is composed primarily of gas that has been expelled from
the galactic disk by supernovae driven winds. This gas must remain available
 as a future reservoir for star formation, even in satellite galaxies.
Finally, we present a simple recipe for the stripping of gas and dark
matter in satellites that can be used in models where subhalo evolution
is not followed in detail.

\end{abstract}




\section {Introduction}
\label{intro}
It has long been known that cluster galaxies are redder, less
active in their star formation,
and of 
earlier type than galaxies in the field (e.g. Oemler 1974; Dressler
1980; Balogh et al. 1997). This
difference has been found to become
 more pronounced towards the centers of clusters
(e.g. Postman \& Geller 1984; Goto et al. 2004; Barkhouse et al. 2009; 
Hansen et al. 2009;
 von der Linden et al. 2009) and perhaps also with
increasing cluster mass (e.g. Weinmann
et al. 2006a; Mart\'{i}nez et al. 2006, Kimm et al. 2009; but see also
De Propris et al. 2004 and Tanaka et al. 2004). 
Understanding these ``environmental effects'' is  not only crucial for
modelling the detailed properties of the global galaxy population, but
it can also help us to investigate fundamental processes of
galaxy evolution like star formation, supernova feedback, feedback by
active galactic nuclei,
and morphological transformations. All of these processes are likely
to be
influenced by the decline of gas accretion in galaxies residing in group and
cluster environments. Clusters and groups of galaxies thus provide unique
laboratories in which we can observe the evolution of galaxies
under conditions
different to those in the field.

Recently, several studies have investigated the fraction of passive
cluster galaxies as a function of cluster-centric radius {\it at fixed
stellar mass}. Most of them find that an increase of the passive
fraction towards the center of cluster is still detectable (Bamford et
al. 2008; von der Linden et al. 2009; this work; but see
also van den Bosch et al. 2008). Investigating the trends at fixed
stellar mass is important since the galaxy stellar mass function
may depend on environment (e.g. Balogh et al. 2001; Baldry et al. 2006;
but see also von der Linden et al. 2009),
and since galaxy properties are strongly correlated with stellar mass
(Kauffmann et al. 2003). In this work, we determine the same
relations in the cluster catalogue of
von der Linden et al. (2007), in which the brightest
cluster galaxies are selected with great care and are thus likely to
mark the approximate center of the cluster. This is crucial for
deriving correct trends as a function of cluster-centric radius.

Semi-analytical models (hereafter referred  to as SAMs, e.g. Kauffmann
et al. 1993; Cole et al. 2000; Croton et al. 2006) track the evolution
of galaxies over time, using simple prescriptions for the treatment of
gas-physical processes  combined with analytical merger  trees of dark
matter haloes or trees derived from N-body simulations.  Most of these
models  treat  environmental  effects  in a  simplistic  manner. 
Motivated by the ``starvation''  scenario suggested by Larson, Tinsley
\&  Caldwell (1980), all  hot gas 
around the  satellites is immediately removed   upon  infall.
The  stripped  gas is  then  made available  for
cooling to the central  galaxy of the corresponding Friends-of-Friends
group.  This simple prescription  leads to satellite galaxies that are
too red  (e.g. Weinmann et  al. 2006b, Wang  et al. 2007)  compared to
observations. This  suggests that  part of the  hot gas  should remain
with the satellite  galaxy. This is not a  minor issue, since satellite 
galaxies constitute a significant fraction of the total galaxy
population and since changing
the prescription for gas stripping in satellites may have a considerable  impact on
the
central galaxy population.  First, satellites eventually merge with
the central galaxies in their halo. If they can grow to higher
stellar  masses, central galaxies  will become  more massive  as well.
Second,  if part  of the  hot gas  stays attached  to  satellites, the
amount  of  gas  available  for  cooling  to  the  central  galaxy  is
reduced. Third, if satellite galaxies merging with the central
  galaxy still contain cold gas, the ensuing star burst will make the
  central
galaxy bluer for a certain period of time, and will result in a higher
final stellar mass.

Attempts to treat environmental effects in SAMs more realistically
have already been made by Kang \& van den Bosch (2008) on the basis of 
the Kang, Jing \& Silk
(2006) SAM  and by  Font et al. (2008), on
the basis  of the  Bower et al.  (2006) model.  Kang \&  van den
Bosch (2008) found that their simple prescription for decreasing 
the efficiency of gas
stripping in satellites leads to a fraction of blue central
galaxies which is higher than observed. They suggested
counterbalancing
this effect  by the inclusion of an additional
prescription for the disruption of satellites.
 Font  et  al.  (2008)  implemented  a more
sophisticated model based on the hydrodynamical simulations of
ram-pressure stripping  by
McCarthy et  al. (2008), and obtained better agreement with observed environmental  
effects than previous semi-analytic models. 

Ram-pressure stripping of the cold disk
gas in satellites has been studied in the SAMs of Okamoto \& Nagashima
(2003) and Lanzoni et al. (2005).  These studies concluded that this effect  has
a negligible impact on the results, because  the complete stripping of
the hot gas halo already makes the  satellites passive.

In this work, we modify the SAM of De Lucia
\& Blaizot (2007, hereafter DLB07) in order to reproduce (i)
the relation
between passive fraction and cluster-centric
radius, as well as the passive fraction in field galaxies 
and (ii) the distribution of specific star
formation rates of satellite galaxies.
 Up to now, this combination of
 observational relations has never been used to
 constrain 
SAMs (but see Diaferio et al. 2001, who study passive fractions as a
function of cluster-centric radius in a set of SAMs implemented on low-resolution
N-body simulations).
We apply a realistic cluster finder to the DLB07 SAM
to allow for a fair 
comparison with observations. We find that a very simple model in
which the diffuse gas halo around satellites is stripped at the
  same rate as the dark matter subhalo loses mass due to tidal
  effects gives good agreement with observations. Interestingly, we
find that this stripping does not proceed exponentially, but linearly,
and that this behaviour is crucial for reproducing the distribution of
the specific star formation rates (SSFR) in satellites. We also implement
a series of modifications  to the SAM, which
lead to improved agreement with the observed passive fractions as a 
function of stellar mass for the central galaxies.
Our goal is not to obtain a full model which
reproduces all the observations; rather, we look at a variety of models
 which allow us to understand better the complex interplay between
 feedback by supernovae (`SN' hereafter), feedback by active galactic
nuclei (`AGN' hereafter) and satellite galaxy stripping.
We also test a model in which the hot gas of satellite
galaxies is removed by ram-pressure stripping, but
find that this model does not reproduce observations in detail.

To summarize, the main goal of this study is to 
provide insight into what causes environmental
effects, on which timescales they act
 and how they might be modelled in SAMs of
galaxy formation.
In section \ref{data}, we present the observational and
semi-analytical data and explain the construction of our mock cluster
catalogue.  In section \ref{observations}, we present
the observational results which we use to compare to models. In
section \ref{sec:discuss_dm}, we discuss the stripping of dark matter subhaloes.
In section \ref{models}, we 
present modified versions of the SAM of DLB07, which we compare
with observations in section \ref{sec:tune}.
Finally, in section \ref{discussion}, we discuss our findings and 
test
a model of ram-pressure stripping for comparison. In
section \ref{summary}, we give a summary of our results.

\section{Data}
\label{data}

\subsection{The Cluster Catalogue}
\label{catalogue}
We use the Cluster Catalogue of von der Linden et al. (2007, hereafter
vdL07) which is based on the SDSS DR4 and the C4 Cluster
Catalogue (Miller et al. 2005). The C4 Cluster Catalogue 
identifies clusters in a parameter space of position,
redshift and colour, making use of the fact that at least a fraction of the
cluster galaxies lie on a tight colour-magnitude relation.
VdL07 carefully identified the brightest cluster galaxies (BCGs) in
these clusters, making sure that galaxies that were not targeted
spectroscopically would not be missed.
 They then
redetermined cluster memberships
and velocity dispersions using the bi-weight estimator by Beers et al. (1990),
and estimated cluster masses from the
velocity dispersions, as described in vdL07.
Their final sample consists of 625 
clusters at redshifts between 0.03 and 0.1, with masses between $10^{12}$
and $10^{15} h^{-1}M_{\odot}$. In most of what follows, we will focus on clusters
with masses $10^{14} - 10^{15}h^{-1}M_{\odot}$ of which there are 214 in the
sample. The 341 clusters with
masses between $10^{13} - 10^{14}h^{-1}M_{\odot}$ will only be used
for comparison.
At the redshift of the most distant cluster, galaxies were observed
down to a limiting magnitude of $M_{r}$ - 5log$h$= -19.75.

In all of what follows, we weight
each galaxy by the inverse of the maximum
volume out to which it can be observed, given the apparent magnitude
limit of the survey. Stellar masses are calculated using the method of
Kauffmann et al. (2003).

\subsection{Central Galaxies}

In  order to obtain  large observational  samples of  central galaxies
needed for  comparison to the SAM, we make use  of the SDSS DR4
group catalogue of  Yang et al. (2007). This  group catalogue has been
constructed  by  applying  the  halo-based  group finder  of  Yang  et
al. (2005) to the New York Value-Added Galaxy Catalogue (NYU-VAGC; see
Blanton  et  al.   2005).   From  this catalogue,  Yang et al. (2007)
  selected  all
galaxies  in  the Main  Galaxy  Sample  with  an extinction  corrected
apparent  magnitude brighter  than $m_{r}=18$,  with redshifts  in the
range $0.01<z<0.20$  and with a  redshift completeness $C_{z}  > 0.7$.
The group catalogue  is publically available and can  be downloaded from
\texttt{http://www.astro.umass.edu/$\sim$xhyang/Group.html}.  We refer
the reader  to Yang et al.  (2007) for a more  detailed description.
  We only use central galaxies from this catalogue
that have redshifts in the range  $0.01<z<0.1$ in this work.

\subsection{The simulation and the SAM} 
We use the SAM by DLB07
which is based on the Millennium Simulation (Springel et
al. 2005). The SAM uses analytical prescriptions for gas
accretion, gas cooling, star formation, SN feedback, AGN
feedback, dynamical friction, merger-induced star bursts
and reionization. These are implemented on the 
merger trees extracted from the Millennium Simulation, which 
follow dark matter haloes and subhaloes over
time. Photometric properties of galaxies are computed using models for
stellar population synthesis and dust. More details can be found in  DLB07, Croton et al. (2006) and references therein. The
SAM has been tuned to reproduce key observations like
the luminosity function at $z$=0 (Croton et al. 2006;
DLB07). The aspect of the
SAM on which we will mainly focus in this paper is the
treatment of gas in satellite galaxies, which will be
explained in more detail in section \ref{standard}.
The cosmological parameters used in this model are in agreement with
the WMAP1 data (Spergel et al. 2003), with $\Omega_{M}$ = 0.25,
$\Omega_{\Lambda}$ = 0.75, $\Omega_{b}$=0.045 and
$\sigma_{8}$=0.9.

\subsection{The SAM Cluster Catalogue}
\label{cluster_catalogue}

We now describe  the construction of our mock  cluster catalogue based
on the  $z$=0 output  of the Millennium  Simulation combined  with the
DLB07  SAM.  Our  goal is  to  mimic the  procedure used  by vdL07  to
determine cluster memberships and velocity dispersions. However, we do
not  try to mimic  the initial  identification of  clusters in  the C4
Cluster Catalogue,  or the method for idenitifying the 
brightest cluster galaxies (BCG).
We make  two assumptions: first, that if  we  impose the same 
magnitude limit on the cluster galaxy sample from 
the Millennium Run simulation  as we do for the vdL07 
catalogue, the resulting sample
of simulated  clusters  can be compared directly with
clusters of the same mass from our observational sample.  
Second, we assume
that  the  BCGs in vdL07 are correctly identified. 
The first assumption is likely not quite correct; 
our sample probably contains more low mass clusters than the
sample of vdL07, since those are more difficult to identify
observationally.  The second assumption seems reasonable. 
VdL07, Best et al.  (2007) and Koester et al. (2007) show that
the BCGs of vdL07 differ systematically from other galaxies of the same mass  and are good
tracers of the cluster centers defined using X-ray images of the hot intracluster gas.
 
Clusters in the Millennium Simulation
are identified with a  Friends-of-Friends (\fof) algorithm (Springel et
al.     2005).  These \fof clusters are not necessarily spherical, but are
often elongated in shape.  
We     select     the     \fof     clusters     with
$10^{13}<M_{200, \rm Mill}/(h^{-1}M_{\odot})<10^{15}$,  with $M_{200, \rm Mill}$  the mass
enclosed  within a sphere  with a  density of  200 times  the critical
density of the universe.   Subhaloes are identified using the \subfind
algorithm (Springel et al. 2001). By
definition, the ``central galaxies'' of clusters 
reside at the center of mass of the
most  massive subhalo  in the  \fof group.  We use  these ``central
galaxies'' as starting points  for determining cluster memberships and
velocity dispersion,  as done in vdL07 for  their observational sample
of BCGs.

We mimic the  observational selection of vdL07 by  only using galaxies
with $M_r  - 5{\rm  log}h <  -18.6$ for the  construction of  our mock
cluster  catalogues,   which  roughly  corresponds   to  the  absolute
magnitude limit  at the median redshift  of the vdL07  sample.  We add
the Hubble flow to the  $z$-direction of the velocities of galaxies in
our simulation box, and we place all galaxies at a minimum distance of
60 Mpc from the virtual observer, so that all clusters are at least as
far away as the nearest cluster in the sample of vdL07. We then follow
the   procedure  explained   in  detail   by  vdL07,   in   which  the
bi-weight estimator of  Beers et al.  (1990) is applied  iteratively to
galaxies surrounding the cluster  center, with the velocity dispersion
and the cluster redshift being redetermined at each iteration step. 
If a  galaxy belongs  to  more than  one  cluster in  the  final list, then 
we follow von der Linden et al. (2009) 
and only count it as a  member of the one  cluster for which
the quantity
\begin{equation}
D_{\rm eff}=\sqrt{(R/R_{200})^2 + (V/\sigma_{\rm 1D})^2}
\label{eq:dist}
\end{equation}
is smaller, with
\begin{equation}
R^2=(x-x_{\rm cl})^2 + (y-y_{\rm cl})^2
\end{equation}
and
\begin{equation}
V^2=(v_{\rm eff, cl} - v_{\rm eff, gal})^2.
\end{equation}

We only deviate from vdL07 in
two ways :

\begin{itemize}
\item
We define two cluster mass bins, one with
$M_{200, \rm Mill}=10^{13}-10^{14}h^{-1}M_{\odot}$ and one with
$M_{200, \rm Mill}=10^{14}-10^{15}h^{-1}M_{\odot}$. If a galaxy lies close to
the centers of two clusters in the same mass bin, we only allow it to
enter into the iterative process for the closer cluster, with the
distance defined by
equation \ref{eq:dist}. This prevents cluster galaxies from being  falsely
linked to a  neighbouring cluster of similar mass. Such a step
is not included in vdL07, where a given galaxy can in principle 
enter into the
iterative process for several nearby cluster centers.
We believe that
this step is justified here, as the density of clusters in the simulation is
higher than in an observational sample. We have checked that
this constraint only increases the fraction of interlopers in the
outskirts of cluster by around 10 \% and does not affect our final
results. 

\item The initial guess for the 1D velocity dispersion is taken to be
  500 km/s for all clusters. VdL07 calculate an initial guess
  individually for each cluster, using the average of the different
  velocity measurements in the C4 Cluster Catalogue. We have checked
  that using an initial guess of 250 km/s or 1000 km/s makes 
  virtually no
  difference to our final results.

\end{itemize}

To enable faster data processing, we only use a subvolume of the Millennium 
Simulation, which encloses around 1/5 of the total simulation box and 
contains the centers of 7586 clusters with masses $10^{13}<M_{200, \rm
Mill}/(h^{-1}M_{\odot})<10^{15}$. 
We have checked that the mass distribution of these
clusters is the same as in the entire simulation box. 412 clusters have a mass
above $10^{14} h^{-1} M_{\odot}$. 
  After applying our bi-weight estimator   
to all of the 7586 central cluster galaxies
 as described above, we remove all clusters which
contain less than 4 members in the final iteration step, as done by vdL07.
We also remove (i) all clusters which
are clearly offset from the relation between velocity dispersion and number
of member galaxies defined by the
bulk of the cluster population and (ii) all clusters for which the
position of the original BCG and the cluster center determined by the
bi-weight estimator of Beers et al. (1990) differ by more than 0.002 in
redshift. This step is necessary since such clusters have undergone
special treatment in vdL07, which we do not repeat here.
It affects less than 3 \% of the clusters in our
sample. Finally, we are left
with 4393 clusters,  constituting  what will be called the ``Beers-Millennium
Cluster Catalogue'' in what follows.  The reason for the high reduction in sample size going from the Millennium Cluster
 Catalogue to the Beers-Millennium Catalogue  is simply that our initial Millennium Cluster sample contains many relatively
low
mass clusters, of which many contain less than four
galaxies. The 1D velocity dispersion of clusters with a
low number of member galaxies also has a large scatter, which makes
convergence of the bi-weight estimator of Beers et al. (1990) less
likely.

We estimate $M_{200}$ for the clusters in our final
sample from the 1D velocity dispersions according to eq. 1 of vdL07.
In all of what follows, $M_{200}$ or the ``cluster mass'' refers to the mass as
obtained in this way.
3151 of the clusters in the ``Beers-Millennium
Cluster Catalogue'' are assigned a
mass above
$10^{13} h^{-1}M_{\odot}$, 747 above 
$10^{14}h^{-1}M_{\odot}$. We therefore have significantly more
galaxies with masses above $10^{14}h^{-1}M_{\odot}$ than in our initial
Millennium Cluster Catalogue. This  
 already indicates that a non-negligible
number of clusters have masses that are overestimated.

\begin{figure}
\centerline{\psfig{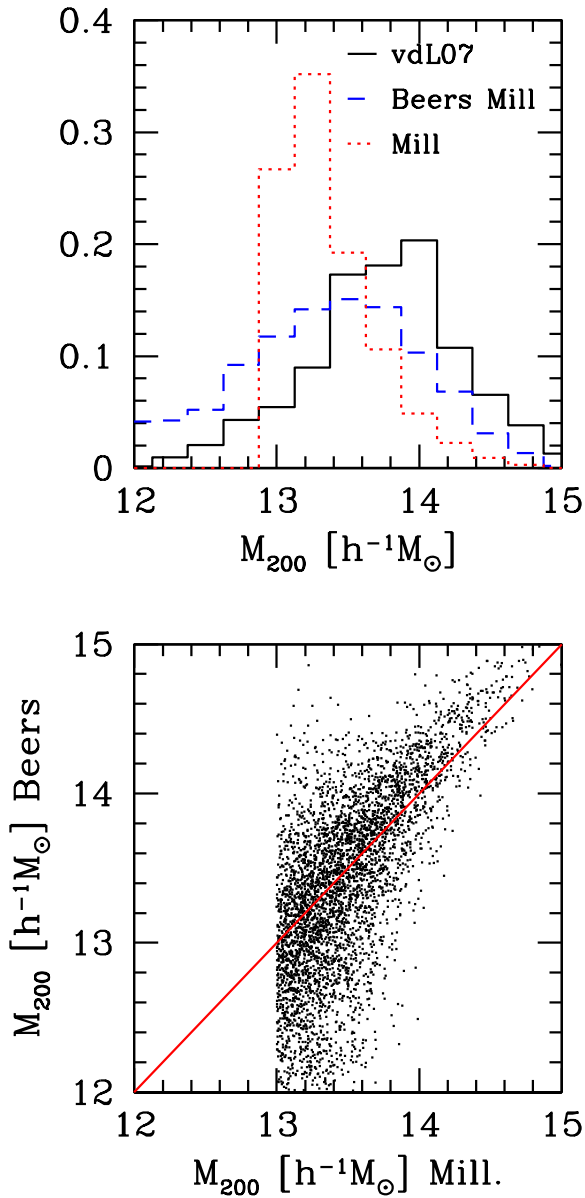}}
\caption{Top panel: The fractional distribution of cluster masses for the 625 clusters in
  the vdL07 catalogue (black line) for the
  4393 clusters in the Beers-Millennium-Catalogue (blue dashed line)
  and for the 7586 clusters in the Millennium Catalogue (red dotted
  lines).
Bottom panel: The relation between true and recovered
  halo masses for the 4393 clusters in the Beers-Millennium
  Catalogue. Clusters with masses around $10^{13}h^{-1}
M_{\odot}$ in the
Millennium Catalogue suffer from a particularly large scatter in the
halo mass, due to their low number of member galaxies which causes a
large spread in 
1D velocity dispersion. }
\label{fig:dist}
\end{figure}

We now compare the properties of the Millennium Cluster Catalogue
catalogue and the Beers-Millennium Cluster Catalogue in detail.
In the top panel of Fig. \ref{fig:dist}, we show the distribution of halo masses in the
three catalogues used in this work: The vdL07 cluster
catalogue, the original Millennium cluster catalogue and the recovered
Beers-Millennium catalogue. 
The original distribution of halo masses is clearly smeared out for the clusters in the
recovered Beers-Millennium catalogue. In the bottom panel of
Fig. \ref{fig:dist}, we show the scatter between the true and
recovered halo masses in the Beers-Millennium catalogue. 
The scatter is large because of the considerable loss of information
that occurs when going from the 3-dimensional to the 1-dimensional
velocity dispersion. This effect is especially severe for low mass clusters
which have only a low number of detectable members.
This means that that a large fraction of 
 the massive Beers-Millennium Clusters, on which we focus our
 analysis,
 have their
masses and radii overestimated, i.e. are low mass clusters in the
Millennium Catalogue. Consequently, the galaxy population
in the cluster outskirts will be strongly contaminated by 
 interlopers. 

\begin{figure}
\centerline{\psfig{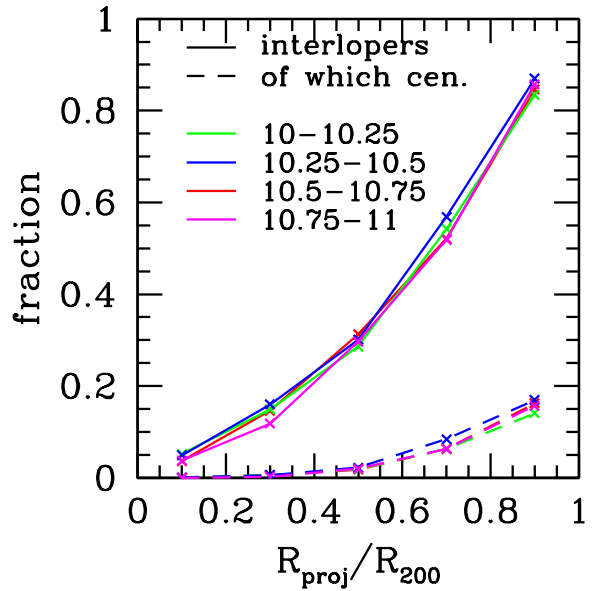}}
\caption{Fraction of interlopers (i.e. galaxies that reside outside
  $R_{200}$ in the Millennium Catalogue) in the Beers-Millennium clusters with masses $10^{14}-10^{15} h^{-1}
M_{\odot}$, in four different logarithmic 
stellar mass bins, as indicated. Note that the definition of an
interloper used here is
very strict and many galaxies classified as interlopers actually belong to the (non-spherical) large-scale
structure associated with the cluster. Dashed
lines show the fraction of interloper galaxies which are centrals in
the Millennium Catalogue.}
\label{fig:interloper}
\end{figure}

We now examine the interloper fraction in our clusters
in more detail. Any galaxy which 
is identified as a member of a Beers-Millennium cluster by the
bi-weight algorithm, but does not
actually reside within $R_{\rm 200, Mill}$ of the corresponding true Millennium
cluster, is defined as an interloper. 
 In Fig. \ref{fig:interloper}, we show the interloper
fractions as a function of cluster-centric radius for clusters with $M_{200}= 10^{14}-10^{15} h^{-1}
M_{\odot}$.  The cluster center is defined to lie at the position of  the central cluster galaxy.
The interloper fraction in the center is very low, but rises steeply towards the
outskirts, as expected. Since we have used a similar method as  vdL07, our
 finding
 indicates that a similar contamination might be present in their
 study\footnote{It is however likely
 that the sample of vdL07 is
   biased towards higher cluster masses, since these are
   observationally easier to detect,
 which could make our contamination slightly higher
 than in vdL07.}. Dashed lines show the fraction of interlopers which are central
galaxies in the Millennium Catalogue. This fraction is very small,
showing
that most of the interlopers are in fact satellite
galaxies; they either belong to the
extended filaments surrounding the cluster (and are as such part of
the \fof cluster), or are
part of infalling nearby groups. Note that as many as 40$\%$ of all satellite galaxies in the
SAM actually reside beyond $R_{\rm 200, Mill}$ of their respective central
galaxy.  These interlopers have by definition
experienced environmental effects in the SAM and cannot be considered
true ``contamination'' from the field; however, the time at which they fell into the cluster 
will differ from galaxies that reside within $R_{200}$ of the cluster center.

We conclude that there is a high interloper fraction in clusters
identified with a typical cluster-finding algorithm, and a relatively 
large scatter
between true and recovered mass. This means that it is crucial to mimic the
observational
cluster finding process when examining environmental effects in models.

\section{Observations}
\label{observations}

\subsection{Defining passive and active galaxies}
\label{sec:def_pass}

\begin{figure}
\centerline{\psfig{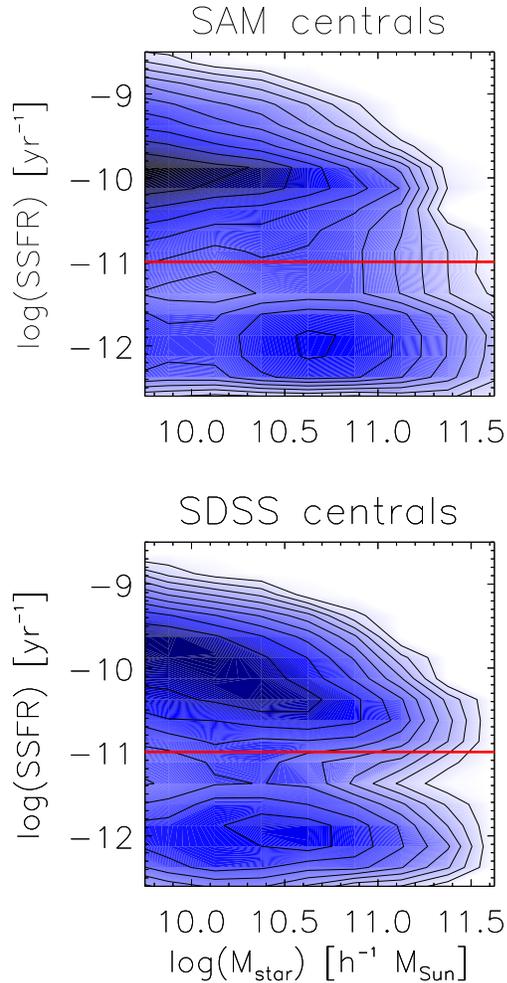}}
\caption{The SSFR as a function
 stellar mass in the SAM and the
SDSS, for central galaxies only. Red lines indicate how we divide
galaxies into ``passive'' and ``active'' in all of what follows.
In the SDSS, we use the Brinchmann et
al. (2003) SSFRs, corrected such that the offset with respect to Salim et
al. (2007) is removed. In the SAM, galaxies with zero star formation rate have
been randomly assigned SSFRs between -11.6 and -12.4, and the same has
been done for SDSS galaxies which have measured SSFRs, but with values 
$<-12.4$, since such low SSFRs tend to have large errors.
In the SDSS, only galaxies
with $z < 0.1$ are used, and a $V_{\rm max}$-correction has been
applied. The number of
galaxies has been normalized to the same value in both panels.}
\label{fig:sca1}
\end{figure}

We use the specific star formation rate (SSFR) to distinguish passive
and active galaxies. The reason for our choice is that the SSFR
 is a physical quantity with a straightforward
meaning. It can be directly taken from the SAM
without any assumptions about stellar population synthesis
models. However, the SSFR is not a directly observed quantity. 

Brinchmann et al. (2003) have derived SSFRs
for galaxies in the SDSS using emission lines, the D4000
spectral index and colours, with colours used for sampling star formation
outside the region covered by the fiber. 
Salim et al. (2007) have redetermined SSFRs 
for a subset of these galaxies from UV and optical photometry, and while
they find good agreement for the ``star-forming'' class of the
Brinchmann et al. (2003) sample, they also find that star formation rates
are  overestimated for the ``star forming low S/N'', the
``Composite'', the ``AGN'' class and the galaxies without measured
H$\alpha$ (see Salim et al. 2007 and Schiminovich et
al. 2007 for more details). 
 We therefore decide to use the SSFR from Brinchmann et al. (2003),
 but with a rough correction, bringing it into statistical agreement
 with Salim et al. (2007). For ``low S/N'' galaxies, ``AGN'' and ``composite''
galaxies, and galaxies with no H$\alpha$, we scale the Brinchmann et
al. (2003) star formation rates down by 0.2, 0.4 and 1.0 dex to correct for   
the offsets seen in  Fig. 3 of  Salim et
al. (2007). The SSFR for the ``star-forming'' galaxies are used
without any correction.

In Fig. \ref{fig:sca1}, top panel, we show the SSFR as a function of
stellar mass for central galaxies in the SAM. A very clear star forming
sequence at around log(SSFR)=-10 can be seen.  
In the bottom panel of the same figure, we show the corrected SSFR
for central galaxies in the SDSS. These were determined using the Yang
et al. (2007) group catalogue and not vdL07,
since it also includes low mass groups and thus provides much better statistics. Every galaxy is
weighted by $1/V_{\rm max}$, where $V_{\rm max}$ is the maximum volume out to which a galaxy 
of that magnitude would be detected in the survey.
We define ``passive galaxies'' as those with
log(SSFR) $<$ -11 both in the SAM and in the SDSS, because this cut corresponds roughly to
the location of the minimum in the bimodal distribution of SSFR in both the model and the observations.

Comparing the overall distributions in the SDSS and the SAM, 
we see that the shape of the star
forming sequence is tilted in the SDSS, unlike in the SAM. This
problem has been noted before (e.g. Somerville et al. 2008), and
seems to be generic to current SAMs (but
see Neistein \& Weinmann 2009)

\subsection{Observational Results}
\label{sec:obs}
Here, we present the observational results that will be used as the basis 
for our model comparisons in what follows.
These include the passive
fraction of central galaxies as a function of stellar mass, the
fraction of passive galaxies as a function of cluster-centric radius,
and the distribution in SSFR for satellite and central galaxies. For
all observations, we apply volume-weighting  to correct for
the apparent magnitude limit of the SDSS.

In our Fig. \ref{fig:altcen}, we show as crosses with errorbars the passive
fraction of central galaxies as a function of stellar mass  
defined using  the group catalogue of Yang
et al. (2007).  As we show in Neistein \& Weinmann (2009), similar results are obtained 
 if the SSFRs of Salim
et al. (2007) are used directly.

We use the  cluster catalogue of vdL07 to  determine passive fractions
as a function of projected cluster-centric radius in the SDSS (crosses
with  errorbars  in Fig.  \ref{fig:gna_b}).  All galaxies within $ 3
\cdot \sigma_{1D}$ of the cluster center are used in this plot.
Errorbars are  determined
using jackknife  resampling.  We have checked that results
are virtually unchanged if we define volume-limited samples that are complete down
to a given limiting stellar mass, 
as done by von der Linden et al. (2009).
For the three lower stellar mass bins, the passive fractions as a function
of cluster-centric radius are nearly identical to what has been found
by
 von der  Linden et
al.  (2009), which is somewhat surprising, since they have defined passive
galaxies
completely differently, with a method based on the light-weighted
  age of the stellar population, rather than the current SSFR as done
  here. This indicates that our way of classifying
galaxies into passive and active is robust, at least at masses below 
$\log(M/h^{-1}M_{\odot})$=10.75. At higher masses, passive fractions
are systematically lower by 10-15 \% than in von der Linden et
al. (2009).

In Fig. \ref{fig:histo_cen} and Fig. \ref{fig:histo_sat} we show as
black histograms the 
distribution of SSFRs for SDSS  central galaxies
as defined by Yang et al. (2007), and for  satellite
galaxies defined here as galaxies located at a distance of less than  $R_{200}$ 
and within $3 \cdot \sigma_{1D}$
from the center of a cluster with mass $\log(M/h^{-1}M_{\odot})>14$ in the
catalogue of vdL07. A clear
bimodality can be seen for both galaxy populations.

\section{Dark matter evolution of subhaloes}
\label{sec:discuss_dm}

\begin{figure*}
\centerline{\psfig{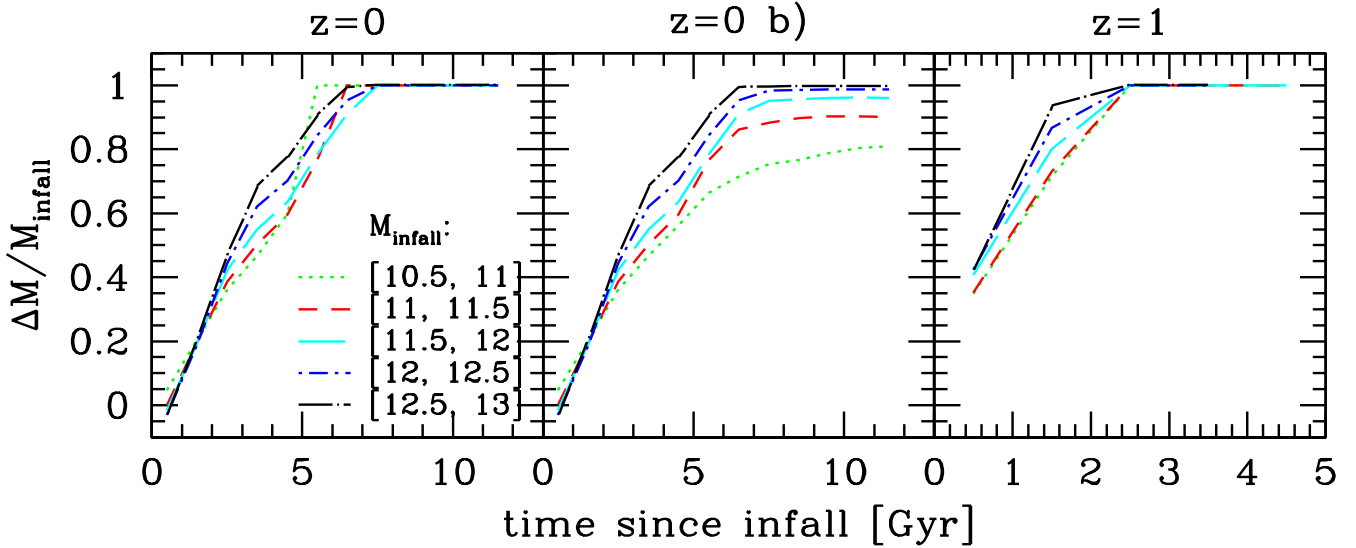}}
\caption{The median fraction of subhalo mass which has been stripped since
  infall as a function of the lookback time to infall. Only subhaloes
  containing galaxies with masses $> 10^{9.5} h^{-1}M_{\odot}$ today in
  DLB07 are used. In the left hand panel, we show results for satellites
  at $z$=0 if the mass of unresolved subhaloes is set to zero. 
In the middle panel, we show results for satellites at $z$=0 if the mass
of unresolved subhaloes is set to the resolution limit of  $1.7 \times 10^{10}
h^{-1}M_{\odot}$, for illustration.
 In the right hand panel, we show results for $z$=1, 
with the mass of unresolved subhaloes set to zero.}
\label{fig:strip_dm}
\end{figure*}

In this section, we investigate the stripping of dark matter subhaloes
in the Millennium simulation. Stripping of subhaloes
 has been studied numerically e.g. by Hayashi
et al. (2003); Gao et al. (2004); De Lucia et al. (2004);
Kazantzidis et al. (2004); Zentner et al. (2005); Giocoli et
al. (2008); Diemand, Kuhlen \& Madau (2007) and 
observationally using weak gravitational lensing signals e.g. by 
Limousin et al. (2007) or Natarajan et al. (2009). The results
presented here will be important for the new prescriptions for satellite evolution that will be introduced in the next section.

In Fig. \ref{fig:strip_dm}, we show the median fraction of dark matter 
which has been stripped since infall, for satellite galaxies in clusters 
and groups at $z$=0 (left hand 
and middle panel) and at $z$=1 (right hand panels), as  a
function of lookback time to infall. We have checked that results do not
strongly depend on host halo mass down to $10^{12}h^{-1}M_{\odot}$.
Different linestyles denote different
bins of $M_{\rm infall}$. Satellite 
galaxies are defined  as galaxies which are part of a \fof group,
but are not the central galaxy of the most massive subhalo. 
For the satellite galaxies at $z$=0, we adopt two different methods
for treating subhaloes which fall below the resolution limit  (sometimes
called ``orphan galaxies'').
In the left hand panel, we assume that their mass falls instantaneously to zero, 
while in the middle panels,  we set set the mass equal to
the effective resolution limit. For the highest mass subhaloes, the results are little 
affected by  resolution and the curves in the  left and
middle panel are similar.
For low mass subhalos, the two prescriptions (not surprisingly) give quite different results.
The plots indicate that the Millennium simulation can be used to follow subhalo 
evolution to high accuracy for systems where $M_{\rm infall}$ is larger than
$10^{11} h^{-1} M_{\odot}$.

Our results for such subhalos indicate that the dark matter 
 is usually nearly completely stripped
after the subhalo  has spent 5-7 Gyr in the cluster.
Interestingly, we find that stripping is surprisingly similar for
subhaloes spanning two orders of magnitude in dark matter mass at
infall, if the dark matter mass of orphan galaxies is set to zero.
We find that for satellite galaxies identified at $z$=0, 
roughly 15-20 \% of the initial dark
matter mass has been stripped per Gyr elapsed since infall.
Note that  the decline is not 
exponential. This means
that \emph{stripping becomes  more efficient the longer a 
subhalo has  been a satellite}. 
This can be explained by the fact that subhaloes which have already spent a significant amount
of time as satellites have  sunk to the center of
the potential well (Gao et al. 2004), where stripping becomes fast, while a large fraction
of subhaloes just having fallen in recently will be at the outskirts
of the \fof group where stripping is inefficient. 
Note that the stripping for an \emph{individual} subhalo typically proceeds 
much less smoothly than the median shown here, as a large fraction of
the stripping
occurs at the first pericenter passage (e.g. Diemand et al. 2007).

For galaxies which are satellites at $z$=1 
(right hand panel of  Fig. \ref{fig:strip_dm}) , the dark matter stripping since
infall
has been significantly more efficient, with $\sim$ 40 \% of the initial
dark matter mass stripped per Gyr in the median. This indicates that the stripping efficiency increases
towards higher
 redshift, which is expected due to the decrease of dynamical
 times. This 
result is 
 in qualitative agreement with the findings of Giocoli 
et al. (2008) and Tinker \& Wetzel (2009).
Note that both Natarajan, De Lucia \& Springel (2007) and Maciejewski et al. (2009)
have found indications that the subhalo masses found with \subfind are systematically
low with respect to the true mass enhancement, or the masses found with more accurate
6D substructure finders. This effect is most severe in the inner regions of
clusters, where the background density of the parent cluster is high (Natarajan et al. 2007).
This means that our estimates of dark matter stripping might be slightly too high,
especially for subhaloes close to the cluster center.
An alternative, and perhaps more accurate, method for estimating the amount of
stripping in subhaloes would be to combine SAMs with analytical
prescriptions based on higher resolution dark matter simulations,
like presented in Taylor \& Babul (2001).

\section{The models}
\label{models}
In this section, we present both the standard SAM of DLB07 and our
modifications of this model.

\subsection{The standard model}
\label{standard}
Here we
give an overview of how gas physics is modelled in central and
satellite galaxies in DLB07 and repeat the basic equations (as
given and explained in more detail in Croton et al. 2006 and
references therein) for clarity.

For central galaxies, gas surrounding galaxies comes in three phases:
The cold gas, the hot gas and the ejected gas.
Gas accreting onto the galaxy from the IGM is added to the
hot phase. From this reservoir, it can cool down to the cold
phase. Cooling is assumed to proceed differently in the ``hot halo
regime'' (where $r_{\rm cool}<R_{\rm vir}$) than in 
the ``rapid cooling regime'' (where  $r_{\rm cool}>R_{\rm vir}$).
In the hot halo regime, cooling rates are given by
\begin{equation}
\label{eq:cool1}
\dot{m}_{\rm cool}=0.5m_{\rm hot}\frac{r_{\rm cool}V_{\rm
      vir}}{R_{\rm vir}^{2}}.
\end{equation} 
$R_{\rm vir}$ is the virial radius of the
dark matter halo, $V_{\rm vir}$ the virial velocity and
$r_{\rm cool}$ is the radius where the
local cooling time $t_{\rm cool}$ is equal to the halo dynamical time.
The local cooling time is given by

\begin{equation}
\label{eq:cool2}
t_{\rm cool}=\frac{3}{2}\frac{m_{\rm p}kT}{\rho_{g}(r)\Lambda(T,
  Z)}
\end{equation}
with $m_{\rm p}$ the mean particle mass, $k$ the Boltzmann
constant, and $\Lambda(T,Z)$ the
cooling function according to Sutherland \& Dopita
(1993) as a function of the temperature of the gas, which is assumed
to be equal to the virial temperature of the halo, and the gas
metallicity.
 $\rho_{g}(r)$ is the hot gas density, which is assumed to have an
isothermal profile
\begin{equation}
\label{eq:iso}
\rho_g(r)=\frac{m_{\rm hot}}{4\pi R_{\rm vir}r^{2}}.
\end{equation}
If $r_{\rm cool}>R_{\rm vir}$, the cooling rate is set roughly to the rate at
which new diffuse gas is added to the halo (see discussion in Croton 
et al. 2006).

The energy input due to radio mode 
AGN feedback according to Croton et al. (2006) is assumed to partially
offset the cooling in high mass haloes, giving rise to a modified
cooling rate
\begin{equation}
\label{eq:cool3}
\dot{m}_{\rm cool, mod}=\dot{m}_{\rm cool} - \frac{L_{\rm
    BH}}{0.5V_{\rm vir}^2}
\end{equation}
with the mechanical heating energy generated by the black hole
accretion $L_{\rm BH}$ given by
\begin{equation}
\label{eq:L_BH}
L_{\rm BH}=\eta  \dot{m}_{\rm BH} c^{2}.
\end{equation}
$\eta$ =0.1 is the standard efficiency with which mass is assumed to
produce energy near the event horizon, and $c$ the speed of light. The
quiescent growth rate of the black hole is given by
\begin{equation}
\label{eq:m_BH}
\dot{m}_{\rm BH}=\kappa_{\rm AGN}
(\frac{m_{\rm BH}}{10^{8}M_{\odot}})
(\frac{f_{\rm hot}}{0.1})
(\frac{V_{\rm vir}}{200 {\rm kms}^{-1}})^{3}
\end{equation}
where $m_{\rm BH}$ is the black hole mass, $f_{\rm hot}$ the fraction of the
total halo mass in the form of hot gas and $\kappa_{\rm AGN}$ a free
parameter
set to $7.5 \cdot 10^{-6}M_{\odot}{\rm yr}^{-1}$ in DLB07.

The cold gas then forms stars according to 
\begin{equation}
\label{eq:stars}
\dot{m}_{\rm star}=\alpha_{\rm SF}(m_{\rm cold} - m_{\rm crit})/t_{\rm
  dyn, disk}
\end{equation}
with $t_{\rm
  dyn, disk}=r_{\rm disk}/V_{\rm vir}$, $r_{\rm
  disk}=3(\lambda/\sqrt{2})R_{\rm vir}$ (motivated by the studies of
Mo et al. 1998 and van den Bergh
et al. 2000), and $\lambda$ the spin
parameter of the dark matter halo in which the galaxy resides.
$\alpha_{\rm SF}$ is a tunable parameter 
and set to 0.03 in DLB07.
The calculation of the critical gas mass follows
Kauffmann (1996) and is based on the observations of Kennicutt (1998)
of a threshold gas density below which stars do not form anymore:
\begin{equation}
\label{eq:crit}
m_{\rm crit}=3.8 \cdot 10^{9}(\frac{V_{\rm vir}}{200 {\rm
    kms^{-1}}})(\frac{r_{\rm disk}}{10 {\rm kpc}})M_{\odot}.
\end{equation}

Supernova
feedback heats cold gas back to the hot phase with an efficiency
directly proportional to the star formation rate:
\begin{equation}
\label{eq:SN}
\Delta{m}_{\rm hot} = \epsilon_{\rm disk} \cdot \Delta{m}_{*}
\end{equation} 
based on the observations by 
Martin (1999), with $\epsilon_{\rm disk} = 3.5$ set by
observational data. Gas in the 
hot phase can be transported to
the ejected phase due to the excess energy present in the hot halo
after reheating, with an efficiency inversely proportional to the
depth of the dark matter halo potential:
\begin{equation}
\label{eq:eject}
\Delta m_{\rm ejected} = (\epsilon_{\rm halo}\frac{V_{\rm
    SN}^2}{V_{\rm vir}^2}-\epsilon_{\rm disk})\Delta m_{*}
\end{equation}
with  $V_{\rm SN}$=630 km ${\rm s}^{-1}$ the mean energy in supernova
ejecta per unit mass of stars formed, based on a standard IMF and
standard supernova theory, and $\epsilon_{\rm halo}$=0.35 tuned to 
reproduce observations. 
Per halo dynamical time, half of the gas in the ejected phase is
assumed to be reaccreted
to the hot phase:
\begin{equation}
\label{eq:reincorporate}
\Delta m_{\rm ejected} = -\gamma \cdot m_{\rm ejected}/t_{\rm dyn}
\end{equation}
with $\gamma$=0.5 and $t_{\rm dyn}$ the halo dynamical time.

\begin{figure}
\centerline{\psfig{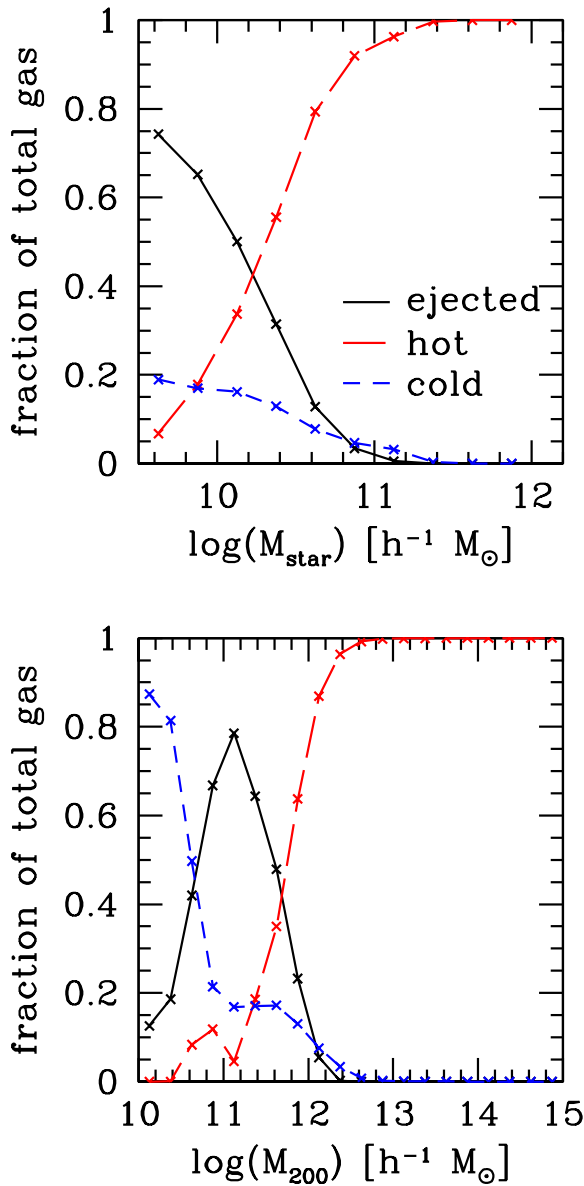}}
\caption{The average fraction of gas in the cold,
  hot and ejected phase, at $z$=0, 
as a function of stellar mass (top panel) and of
  halo mass (bottom panel), for central galaxies with stellar masses above
  log($M/h^{-1}M_{\odot}$)=9.5 in the SAM of DLB07.}
\label{fig:gas}
\end{figure}

In Fig. \ref{fig:gas}, we plot the fraction of the total gas in the
cold, hot and ejected phase in central galaxies at $z$=0 
both as a function of galaxy stellar mass
and of dark matter halo mass in the DLB07 SAM. Clearly, the ejected phase strongly
dominates both at low stellar and halo masses, where the hot phase is relatively
unimportant. This is a consequence of the rapid cooling and the very 
efficient SN
feedback in low mass haloes, and indicates that the ejected phase, and
how it is reaccreted, is of crucial importance for the
evolution of 
low mass galaxies (see also Oppenheimer et al. 2009).

Satellite galaxies in DLB07 are defined as all galaxies which are
member of \fof groups, but not the central galaxy of the most massive
subhalo. In all our models presented below, we will follow this
definition.
If a galaxy becomes a satellite in DLB07, both the hot and ejected gas is
removed. Any hot and ejected gas produced after infall is stripped
as well. Due to the strong efficiency of SN feedback in the model,
this leads to a quick depletion of cold gas and to the cessation of
star formation. All stripped satellite gas is added to the hot gas of
the central galaxy.

\subsection{Changes to the standard model}
\label{sec:alternative}
In what follows, we describe our basic changes to the standard DLB07, 
which are ingredients of all our models 1) - 4) described below.

\begin{itemize}
\item We allow cooling from the hot to the cold gas, and
  reincorporation from the ejected to the hot gas, for satellite
  galaxies. More details are given below.
\item In the DLB07 SAM, processes like cooling and star formation in galaxies
 are 
  correlated with the properties of the corresponding halo, as
  apparent from the equations given above. 
After a galaxy
  becomes a satellite, this is not necessarily appropriate. For
  example, it 
  might not generally be true that the disk radius of a satellite decreases in
  proportion to the radius of its dark matter halo, as assumed in
  DLB07. De Lucia \& Helmi (2008) therefore fix the disk radius
  of the satellite at infall. It is also 
unclear how the hot gas halo redistributes itself
  after part of the dark matter has been stripped.
 In our modified model, we  
fix $R_{\rm vir}$, $\lambda$ and $V_{\rm vir}$ for satellites
 in eq. \ref{eq:cool1}-\ref{eq:reincorporate} 
at infall and do not let it evolve anymore. This means that star formation,
cooling, ejection and reincorporation for satellites are calculated
according to the subhalo properties at infall.
We still follow the true evolution of the dark matter subhalo
in detail, as it is used for (i) determining dynamical friction
timescales and (ii) our new recipes for satellite stripping, as
described below.

\item Unlike DLB07, we allow AGN
  feedback for satellite galaxies as
  described in eq. \ref{eq:cool3}, \ref{eq:L_BH} and \ref{eq:m_BH} in
  our models 1) - 4), again using $R_{\rm vir}$ and $V_{\rm vir}$ as
  determined at infall.
  As radio AGN activity has been observed in
  satellite galaxies (e.g. Best et al. 2007; Pasquali et
  al. 2009a), this step seems well justified.

\item It has been found in previous work (e.g. Weinmann et al. 2006b)
  that satellite galaxies in DLB07 are too red. As mentioned before,
  DLB07 assumes that all hot and ejected gas is removed from
  satellites.
We find 
  that letting satellite galaxies keep their entire hot gas, while 
  completely stripping the ejected phase,
  still produces too many passive low mass satellites, as only a small
  fraction of the gas in these galaxies is actually in the hot mode 
(see Fig. \ref{fig:gas}). We therefore use a different approach. 
At each timestep, we assume that some fraction $f_{\rm strip}$ of 
$M_{\rm diffuse}=M_{\rm hot}+M_{\rm ejected}$ is stripped. 
We assume that the ejected gas is always stripped first, which
means that the hot gas reservoir is only stripped when 
$f_{\rm strip} \cdot M_{\rm diffuse} > M_{\rm ejected}$. 
All the stripped gas is always added to the hot gas of the central
galaxy in the corresponding \fof group. 
In our models 1) - 3), we strip the diffuse gas halo
of satellite galaxies  in proportion to the dark  matter subhalo. 
This
means that if a given satellite galaxy loses a fraction of its dark
matter mass between two subsequent timesteps $i$-1 and $i$, the same
fraction 
$f_{\rm strip, i}$ of the
current mass in diffuse gas is removed. 
\begin{equation}
f_{{\rm strip}, i}= 1-\frac{M_{{\rm DM}, i}}{M_{{\rm DM}, i-1}}
\end{equation}
To calculate cooling rates, we need to make an assumption about the
hot gas density profile of satellites. Here we simply assume that the
 the remaining
 hot gas redistributes itself according to
 eq. \ref{eq:iso} (with $R_{\rm vir}$ as recorded at infall) after
 each stripping event. This means that the gas becomes progressively
 more diluted, which additionally decreases cooling rates.
This method differs from Font et
al. (2008), who determine a stripping radius
and remove all gas outside of this radius, causing  the hot gas halo
to become increasingly compact. We think  that our
simpler approach is justified, because the
distribution of the hot gas  in satellite galaxies 
is highly uncertain;  continued
energy input by SN feedback might well cause  the hot gas halo 
to expand  after a
stripping event has occurred. Also, is is not entirely clear how
$R_{\rm vir}$ and $V_{\rm vir}$ should be calculated for
subhaloes.

Finally, we note that if the satellite galaxy falls below the resolution limit,
 all the hot and ejected gas is
stripped at the timestep of transition. After that, the gas content of the galaxy
only evolves as a result of star formation and SN feedback.

\end{itemize}

\subsection{Specific Modifications}
We carry out a series of progressive modifications to the semi-analytic
models of DLB07,
denoted modifications 0) - 3), 
as described below. We also try out a simpler approach, which does not
follow the stripping of the DM haloes in detail, denoted modification 4).

\begin{itemize}

\item 
In our modification 0), the only change with respect to DLB07 is that we use
$R_{\rm vir}$ and $V_{\rm vir}$ as recorded at infall
in eq. \ref{eq:cool1}-\ref{eq:reincorporate}, as described in section
\ref{sec:alternative}, second point. Since all hot and ejected gas is
still removed from satellites, cooling and reincorporation rates for
satellites are
still zero, as in DLB07.
We have checked that results are virtually indistinguishable from
DLB07, with the only exception of satellite morphologies, which are
brought into better agreement with observations. We plan to study
this issue in future work.

\item 
In our modification 1), we implement all the
changes described in section \ref{sec:alternative}, i.e. we strip the
diffuse gas in proportion to the dark matter. 

\item 
For modification 2), we additionally decrease
the quiescent hot gas accretion rate to the black hole
 $\kappa_{\rm AGN}$ (as defined in eq. \ref{eq:m_BH}) by a factor
of 5, and
 we let satellite galaxies
residing in host haloes with masses below $10^{12}
h^{-1}M_{\odot}$ keep all their hot and ejected gas.
 The motivation for this is discussed  in section
\ref{sec:model2}. For illustration, we also present a modification 2b), 
which only includes the decrease in $\kappa_{\rm AGN}$.

\item  
While the dark matter subhalo mass 
for most simulated satellite galaxies continuously
decreases, there are also cases in which a subhalo seems to experience
dark matter growth. The reason for this is mainly that the
\subfind algorithm (Springel et al. 2001), which is used to identify
subhaloes, defines the boundary of a subhalo as the point where the
density is equal to the background density. The same subhalo will
therefore have a lower mass assigned if it is close to the center of
its host halo, compared to when it is residing in the outskirts. This
can lead to artificial mass fluctuations and thus to overstripping of
diffuse
gas in our approach. We correct for this effect
in our modification 3).
If the dark matter subhalo
mass at a given timestep is higher than in the previous timestep, we
add the appropriate amount of gas to the hot gas halo of the satellite. 

\item 
Modification 4) differs from the previous ones, since it does not rely
on following the stripping of dark matter subhaloes. This
prescription may be used by SAMs which do
not follow the stripping and the orbits of subhaloes after they have
become part of a \fof group (like e.g. Somerville et al. 2008).
We simply store the mass in diffuse gas at the
time of infall, and strip some fixed fraction
 of this initial (not the current!)
diffuse gas mass per Gyr. We determine this fraction by tuning
the model to fit observations. We also require that the 
median stripped dark matter fraction roughly agrees with
the results calculated directly from the simulations, which have been discussed in  section 
\ref{sec:discuss_dm}. We find that a good solution is to strip 
20 \% of the initial gas mass per Gyr
if the subhalo mass at infall is $<5 \times
10^{11} h^{-1}M_{\odot}$, and  10 \% per Gyr if it is above this
mass. As in our models 2) and 3), we decrease the AGN feedback
efficiency, satellites residing in
groups with masses below $10^{12} h^{-1}M_{\odot}$ are not stripped at all, and as in models 1) - 3),
we always first strip the ejected, and then the hot gas.

\end{itemize}

A summary of our models is given in Table 1.
\section{Results}
\label{sec:tune}
In this section, we compare the model results
 to the observational results described in section \ref{sec:obs}.
For  all plots  in this  section, we  use the same subvolume  of the
Millennium Simulation as used in section \ref{cluster_catalogue}. 
The bi-weight algorithm described in the same section is re-applied
for each new model.  
For the model results shown in the plots in this section,
central galaxies are defined as the galaxy belonging to the most
massive subhalo in a given \fof group, while satellites are defined to be those galaxies
located at a projected distance of less than  $R_{200}$ and within 
$3 \cdot \sigma_{1D}$ from the center of a  cluster with
log$(M/h^{-1}M_{\odot})>14$ (with $M$ determined using the
bi-weight algorithm).

\begin{table}
\label{the_models}
\caption{Overview of the five modified version of DLB07 discussed in this paper. 
Modifications 1) - 4) include all the changes discussed in
section \ref{sec:alternative}, modification 0) only the first point.
 $\kappa_{\rm AGN}$ is the  the quiescent
hot gas accretion rate to the black hole in $M_\odot/{\rm yr}$, as
defined in Croton et al. (2006). For modification 4), $f_{1}$ and  $f_{2}$ are the fractions
  of the initial diffuse gas stripped per Gyr. $f_{1}$ is for an
  initial subhalo mass of $< 5 \cdot 10^{11}h^{-1}M_{\odot}$, $f_{2}$ for higher
  masses.}

\begin{center}
\begin{tabular}{|lllll|}
\hline
& $\kappa_{\rm AGN}$ & sat. stripping &  line  \\

\hline\hline
0) & 7.5 $\cdot 10^{-6}$ & instantaneous &  blue dotted\\
1) & 7.5 $\cdot 10^{-6}$  & $\propto$ DM & 
cyan solid\\
2) & 1.5 $\cdot 10^{-6}$ & $\propto$ DM $> 10^{12}h^{-1}M_{\odot}$ &  green long-dashed\\
3) & 1.5 $\cdot 10^{-6}$ & as 2) + accretion & red dashed\\
4) & 1.5 $\cdot 10^{-6}$ & $f_{1}$=0.2, $f_{2}$=0.1& magenta dot-dashed\\
\hline
\end{tabular}
\end{center}
\end{table}
\subsection{Modification 0}

Implementing modification 0) gives results that are
 virtually indistinguishable from
DLB07.
In Fig. \ref{fig:altcen}, we compare the passive fraction of central
galaxies in modification 0) (blue dotted line) with observations (data
points). Surprisingly, the agreement is rather poor. At intermediate stellar
masses, the passive fraction is underproduced, while it is
overproduced at high stellar masses. 
The disagreement at intermediate stellar masses might partially be due to the
contamination of the central galaxy sample of Yang et al. (2007) by
satellite galaxies\footnote{The fraction of galaxies which are
  satellites at masses log$(M_{\rm star}/h^{-1}M_{\odot}) \sim$ 10 in
is higher by $\sim$ 15 \% in the
SAM than in Yang et al. (2007). Under the rather extreme assumptions
that the satellite fraction in the SAM is correct, 
 that these additional 15 \% of
satellites are wrongly counted as centrals 
in the Yang et al. (2007) group catalogue and that all of them are
passive, we would be able to roughly account
for the observed difference.}. 
However, it is more likely that DLB07 underproduces the 
number of passive central galaxies at
intermediate masses. This problem has already been noted by Baldry et
al. (2006) and Bolzanella et al. (2009) who found that Croton et
al. (2006) model did not produce enough blue central/field galaxies.
The apparent overproduction of passive galaxies with low SSFR  at high stellar masses, on the
other hand, is more puzzling, since the fraction of {\it red}, massive
galaxies seems to be in agreement with observations 
(e.g. Weinmann et al. 2006b). There are two potential explanations for
this apparent discrepancy.
First, precise determination the SSFRs  of massive, red galaxies
is difficult, and it is entirely possible that they may have been overestimated in our data. 
We note that  von der Linden et al. (2009) find higher passive
fractions at  log$(M_{\rm star}/h^{-1}M_{\odot}) \sim$ 10.7 than we do
here.  Second, it is important to remember that passive fractions
defined according to colour are different from those defined according
to SSFRs.  Colours are sensitive to dust attenuation as well
as the fraction of young stars in the galaxy, so the agreement 
with the fraction of
red objects found in previous work (e.g. Croton et al. 2006) 
may just have been fortuitous.
The  main goal of the current  paper is to improve the treatment of
environmental effects, so we will not investigate this issue in detail;
we simply decrease AGN feedback in some of 
our models, as described below.

In Fig. \ref{fig:gna_b}, we compare modification 0) (blue dotted line) with the passive fractions as
a function of cluster-centric radius. Clearly, the model overproduces
passive satellites, in agreement with previous findings (e.g. Weinmann et
al. 2006b). In Fig. \ref{fig:histo_cen}
and Fig. \ref{fig:histo_sat}, we show the distribution of SSFR for central
and satellite galaxies in modification 0) (blue lines) compared to 
observations (black lines).  
The number of satellite galaxies with high star formation rates is
clearly too
low compared to the observations.
For the central
galaxies, the location of the
active peak is not correctly reproduced,  as a result  of  the missing tilt in
the relation between SSFR and stellar mass discussed before. For
all models discussed below, the location of the two peaks  for the
central galaxies is very similar,
and only the relative heights change. In what follows, we will
therefore 
simply concentrate on
the passive fractions for the central galaxies.

\subsection{Modification 1}
In our modification 1), we strip the
diffuse gas at the same rate as the dark matter of subhaloes is
stripped due to tidal effects. 
We find that the overproduction of
passive central galaxies get slightly worse (see
Fig. \ref{fig:altcen}, solid cyan lines). The reason for
this is that less hot gas is now available for cooling to the central
galaxies. For the satellite galaxies, results are improved with
respect to DLB07 in the two lower mass bins, while results are similar
in the two higher mass bins (see Fig. \ref{fig:gna_b}, solid cyan
lines). In Fig. \ref{fig:histo_sat}, cyan lines indicate the distribution of SSFRs
for satellite galaxies. 

\begin{figure*}
\centerline{\psfig{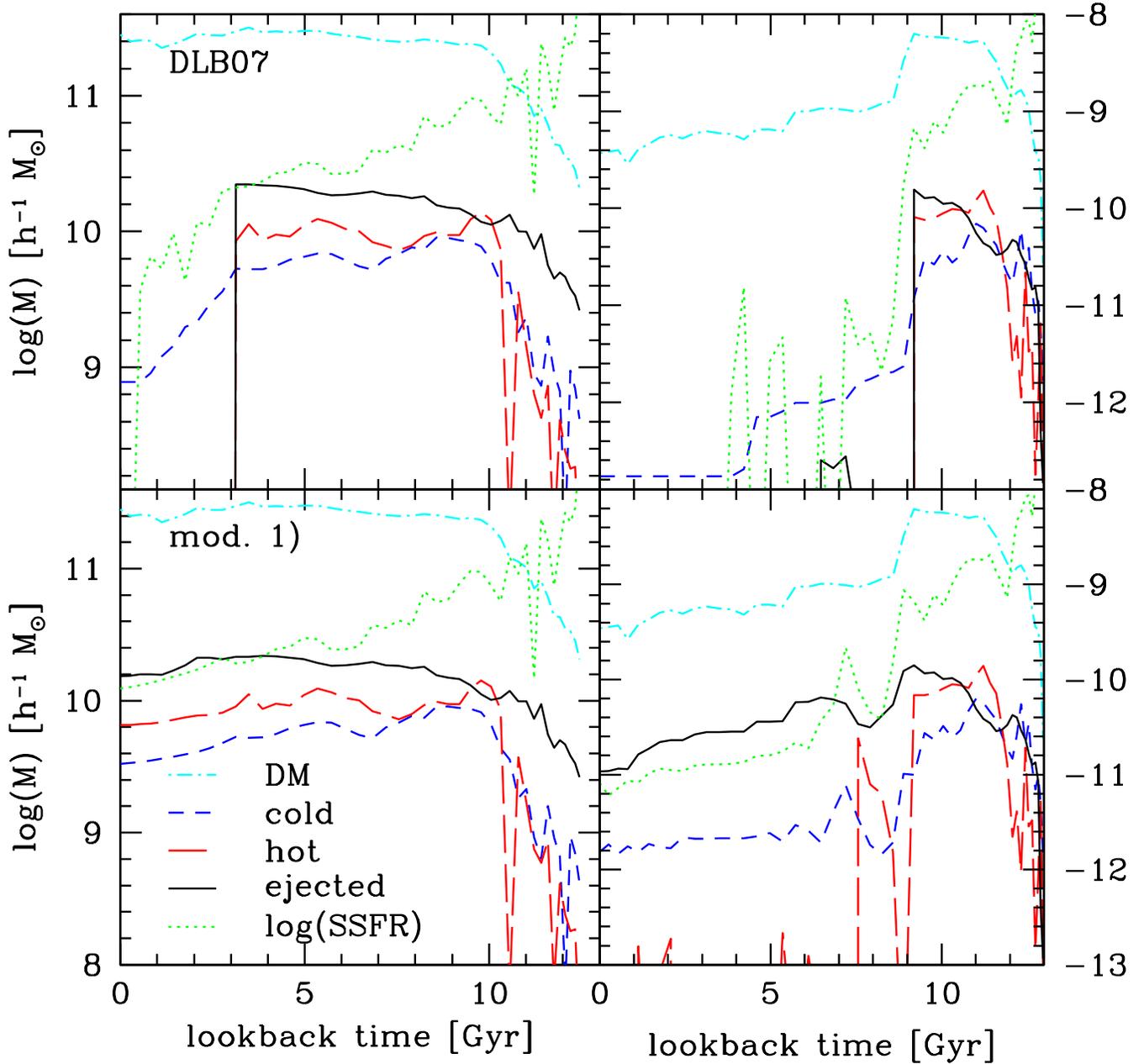}}
\caption{Here we show the histories of two random galaxies with
  $\log(M_{\rm star}/h^{-1}M_{\odot}) \sim 10$, for the DLB07 model (top
  panels) and in our modification 1) (bottom panels). We plot the
  masses in cold, hot and ejected gas and of the DM subhalo,
 and log(SSFR) (with values of
  the latter given
at the right axis of the right hand plots). For both example galaxies, 
the hot and ejected gas mass drops to zero after infall in DLB07,
which happens around 3 Gyr before present in the galaxy shown in the left hand
panels, and at around 9 Gyr before present in the galaxy shown in the
right hand panels. The galaxy in the right hand panel shortly becomes
a central again at around 7 Gyr. At $z$=0, 
the galaxy in the left hand panel has a stellar mass of 
$\log(M/h^{-1}M_{\odot})$=10.1 (10.07) in our
modification 1) (in DLB07). The galaxy on the right hand panel has a
mass
of $\log(M/h^{-1}M_{\odot})$ 10.18 (10.09).
 }
\label{fig:example}
\end{figure*}

For illustration, we compare the star formation, dark matter and gas accretion histories of
two random satellite  galaxies at $z$=0 with $\log(M_{\rm
  star}/h^{-1}M_{\odot}) \sim 10$. Results are shown for the DLB07 prescriptions (top panels) and
for modification 1) (bottom panels) in
Fig. \ref{fig:example}. In the left hand (right
hand) panels, 
the galaxy becomes a satellite at around a lookback time of 3 Gyr (9 Gyr), 
which causes the hot and ejected gas reservoir to fall to zero in
DLB07. In our
modification 1), stripping is much slower. Despite the fact that we always strip the
ejected gas first, the hot gas reservoir is depleted more quickly for the 
galaxy in the right hand panel. This is  due to the fact that
nearly all hot gas that is added by 
SN feedback or that is re-incorporated from the ejected phase  cools down efficiently in
low mass galaxies.

\begin{figure}
\centerline{\psfig{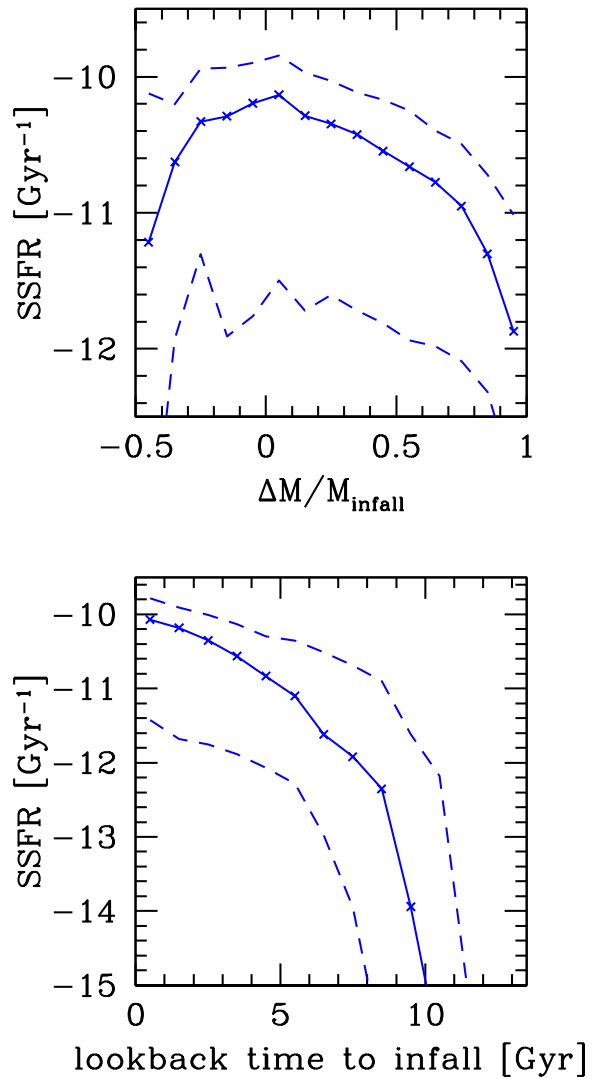}}
\caption{
We show the median SSFR and the region where
68 \% of the values lie, as a function both of the fraction of mass
which has been stripped since infall (top panel) and as a function of
the lookback time to the last time the galaxy became a satellite
(bottom panel). Results here are shown for galaxies with masses of
  $\log(M/h^{-1}M_{\odot})$=10-10.5, at $z$=0, in our modification 1).
}
\label{fig:strip_dm2}
\end{figure}

In Fig. \ref{fig:strip_dm2}, we show the relation between 
the median SSFR and the fraction of the dark matter subhalo mass 
stripped since infall (top panel), and the lookback time to infall
(bottom panel) in modification 1) for galaxies with stellar masses
$\log(M/h^{-1}M_{\odot})=10-10.5$.
It can be seen that galaxies typically only become
passive
after around 80 \% of the dark matter subhalo (and thus of the diffuse
gas) has been stripped, and only after around 5 Gyr since infall.

\subsection{Modification 2 and 2b}
\label{sec:model2}
We try to address the overproduction of passive
central galaxies 
caused by  modification 1)  by
lowering 
 $\kappa_{\rm AGN}$, as defined in eq. \ref{eq:m_BH}. 
Green dot-long dashed lines in  Fig. \ref{fig:altcen} 
show the effect of modification 2b), in which 
$\kappa_{\rm AGN}$ is lowered by a factor of 5. The satellite
stripping is treated exactly as in model 1).
 While results are improved at the
massive end, the passive fractions fall well below the observations 
at intermediate stellar masses. 
After some experimenting, we found that this problem can largely be solved  if
 satellite galaxies
in host haloes with masses below $10^{12}
h^{-1}M_{\odot}$ keep all their hot and ejected gas.
 The reason  is that this  substantially reduces the
amount of gas available for cooling to the central galaxy in such haloes.
We implement this change in our modification 2), shown 
as green dashed lines in Fig. \ref{fig:altcen}. Agreement with
observations is improved compared to DLB07, although it is still not perfect. 

Why should gas stripping cease to be effective in halos with masses below 
 $10^{12} h^{-1}M_{\odot}$?
One reason is  that a hot gas halos
no longer form at  these halo masses (see Fig. \ref{fig:gas}), which means that 
ram-pressure stripping will no longer be effective.
Our results are in agreement  with McGee et al. (2009), who 
find that observations are 
best reproduced under the assumption that  no significant
quenching of star formation occurs in host haloes with masses below $ \sim
10^{12} h^{-1}M_{\odot}$. 
We note that significant dark matter stripping does still occur in  
halos with masses below $10^{12} h^{-1}M_{\odot}$.
Our  results should thus be taken as an indication  that the prescription
for gas stripping that we have introduced, in which the gas is removed
at the same rate as the surrounding halo loses mass, is simply an {\em approximation} that
appears to give roughly the right answer for satellite galaxies in massive halos.
It should not be regarded as a  correct representation of the physics 
that controls the stripping of the gas.  

Modification 2) is quite successful in reproducing passive
fractions as a function of cluster-centric radius (see
Fig. \ref{fig:gna_b}, green long-dashed lines). The remaining disgreement is caused by
the fact that massive central galaxies are still slightly too passive in this
model --   this problem propagates to the
satellite galaxies irrespective of
the stripping that we implement.
  However, if we
look at the detailed distribution of the SSFR of satellite galaxies
(Fig. \ref{fig:histo_sat}, green histogram), we find that the star forming 
peak of the 
satellite galaxies is somewhat too low, and that that there are too
many satellite  galaxies with intermediate SSFR present. 

\subsection{Modification 3} 
In  modification 3), we correct for fluctuations in
subhalo masses, by letting satellite galaxies ``accrete'' gas
from the hot gas
reservoir of their host halos, if their dark matter subhalo experiences
apparent mass growth. 
While we find that the distribution of SSFR
of satellites
is now better  reproduced  
(Fig. \ref{fig:histo_sat},
red histogram), 
the  passive fractions (Fig. \ref{fig:gna_b}, red dashed lines) do not agree as well with the data.\\

\subsection{Modification 4}
\begin{figure}
\centerline{\psfig{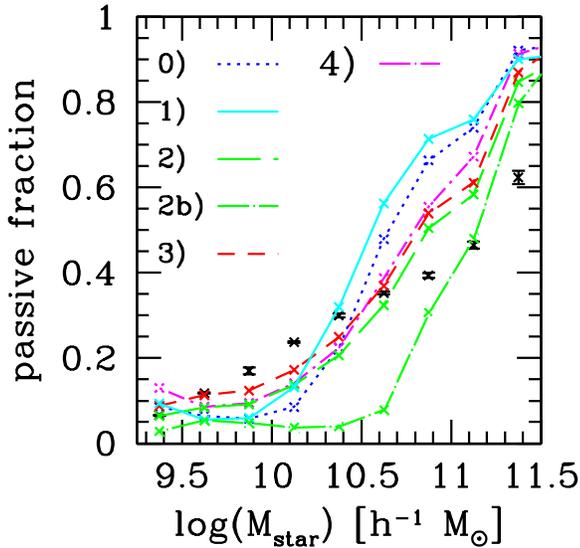}}
\caption{Here we show the impact of our different modifications
  to the SAM on the passive
  fraction of central galaxies. Different linestyles
  correspond to different modifications, as
  indicated and described in the text. Crosses with errorbars show the
observational results.}
\label{fig:altcen}
\end{figure}

\begin{figure*}
\centerline{\psfig{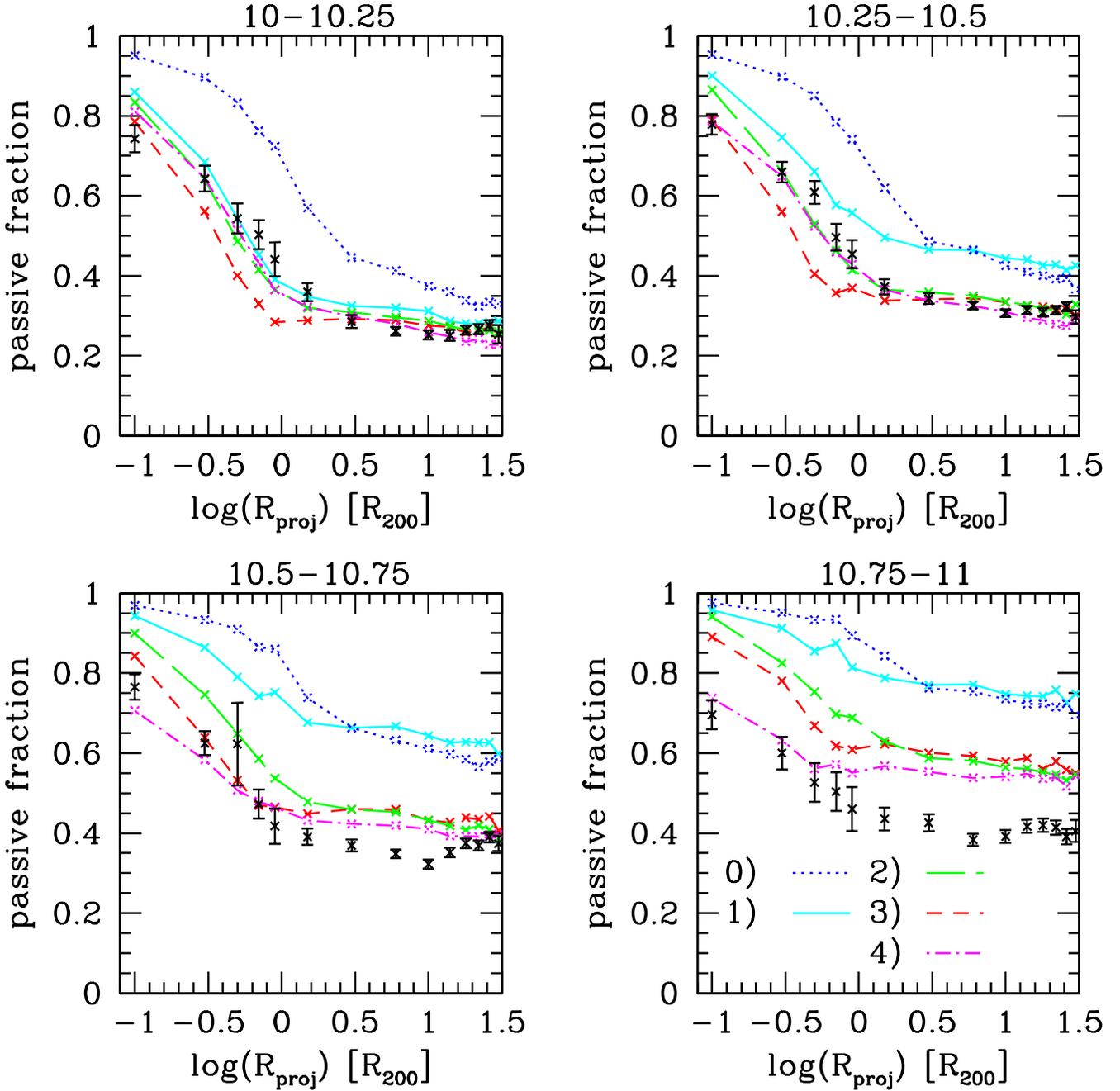}}
\caption{Here we show the impact on our modifications 
on the passive
  fraction as a function of cluster-centric radius in four different
  stellar mass bins, as indicated on the top of the panel. Different linestyles
  correspond to four alternative SAMs, as
  indicated and described in the text. Crosses with errorbars show the
observational results. Only clusters with masses above
$10^{14}h^{-1}M_{\odot}$
are used.}
\label{fig:gna_b}
\end{figure*}

\begin{figure}
\centerline{\psfig{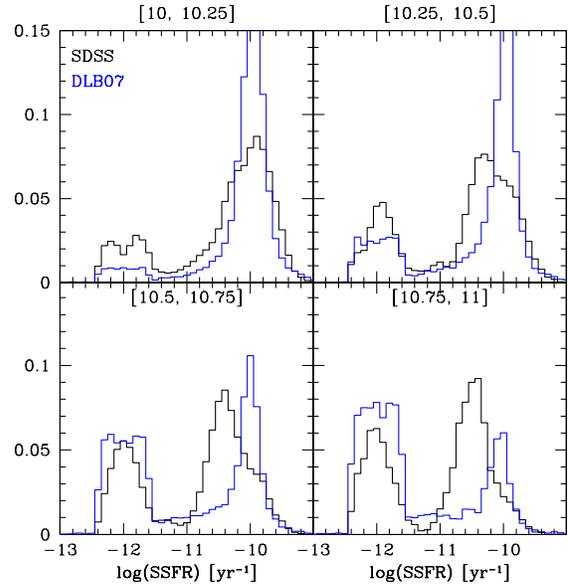}}
\caption{The distribution of the central SSFR in DLB07
 compared to the SDSS central galaxies according to Yang et
  al. (2007).
Galaxies with zero SSFR in the SAM are randomly
distributed between log(SSFR)=-11.6 and log(SSFR)=-12.4, and the same is done
for SDSS galaxies with measured log(SSFR) $<$ -12.4. Results for our
alternative SAMs are not shown since the location of the peak is
identical
with DLB07, and only the relative peak heights change,
which is
information already given in Fig. \ref{fig:altcen}.}
\label{fig:histo_cen}
\end{figure}

\begin{figure*}
\centerline{\psfig{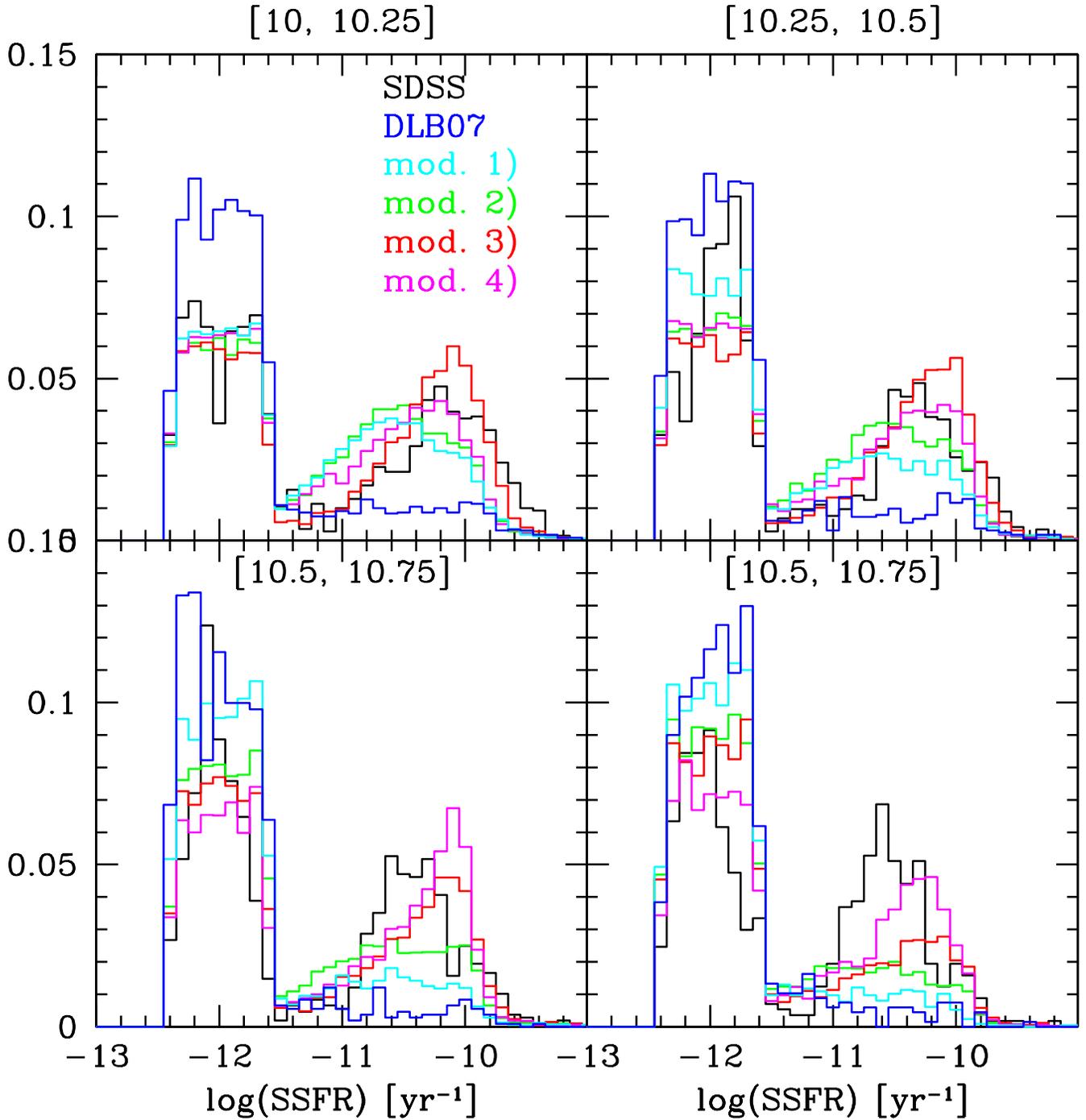}}
\caption{The distribution of the satellite SSFR in DLB07 and our 4
  modifications 
 compared to observed satellite SSFRs.
Galaxies with zero SSFR in the SAM are randomly
distributed between log(SSFR)=-11.6 and log(SSFR)=-12.4, and the same is done
for SDSS galaxies with measured log(SSFR) $<$ -12.4. Both in
  the models and the observations, satellite galaxies are defined as
  galaxies residing within $R_{200}$ from the centers of clusters with
  masses $> 10^{14}h^{-1}M_{\odot}$ as
  found by the bi-weight algorithm described in section \ref{cluster_catalogue}.}
\label{fig:histo_sat}
\end{figure*}

By construction, modification 4) produces results that are in good agreement 
with observations, both in terms
of the distribution of SSFR and the passive fractions as a
function of cluster-centric radius. The results 
are  shown in Fig. \ref{fig:altcen} -- \ref{fig:histo_sat} as magenta
dot-dashed lines.
Note
that it is not possible to obtain a similar level of agreement
if we assume that gas is stripped exponentially. While it is
relatively easy to reproduce passive fractions, we find 
that an exponential decline
of the gas reservoir cannot at the same time produce a
bimodal SSFR distribution for the satellites. Also Balogh et 
al. (2009) have pointed out that reproducing the detailed 
bimodal colour
distribution of satellite galaxies with SAMs is not trivial.

\subsection{The impact on the global properties of the galaxy
  population}

In this section, we investigate the properties of the global galaxy
population
in our alternative models. We have not tuned our models to match these
and it would  thus not be surprising if the models were not to  agree
with the observations.
Here, the full simulation box is used. Note that for all
results shown here, there is no visible difference between DLB07 and
our model 0).

\begin{figure}
\centerline{\psfig{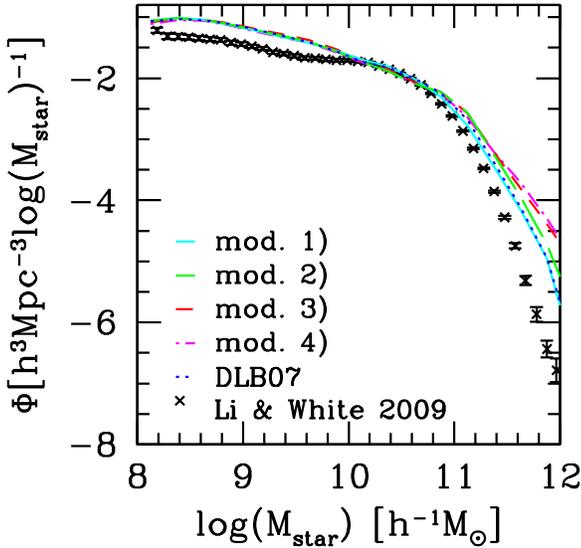}}
\caption{The stellar mass functions in the standard model of DLB07
(blue dotted lines) and in our modifications of this model, as 
indicated.
 The observed mass function is a modified version of the one
presented in Li \& White (2009), corrected according to Guo et al. (2009).}
\label{fig:mf}
\end{figure}

The stellar mass functions in the different models 
are compared to the observations of 
Li \& White (2009), corrected according to Guo et al. (2009), in the
top panel of Fig. \ref{fig:mf}. All models overproduce the massive end
of the stellar mass function. In our models 2) - 4), this problem is accentuated. This is due to the decreased
efficiency of AGN feedback that we need in order to reproduce the
fraction of passive central galaxies. This clearly indicates that 
simply decreasing AGN feedback is not a viable solution; maybe, 
it would be  more appropriate to make AGN feedback more stochastic, for
example by shutting down gas cooling after major mergers (e.g. Hopkins
et al. 2008). Interestingly, the overproduction of luminous galaxies is
more severe in our modification 3) than in our modification 2), which only differ in
details of the implementation of satellite gas stripping. This 
indicates that the treatment of satellite can have a significant impact on
the general galaxy population.

\begin{figure}
\centerline{\psfig{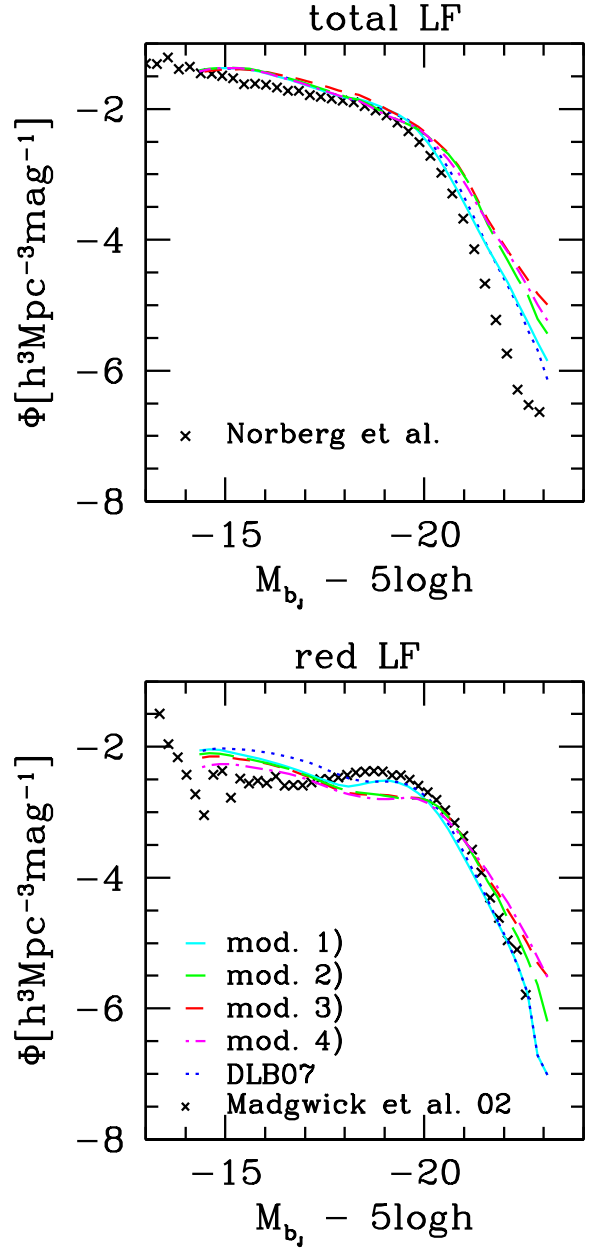}}
\caption{The total $b_{J}$-band luminosity
  function (top panel) and the $b_{J}$-luminosity function of red
  galaxies alone (bottom
  panel) for the standard model of DLB07
 (blue dotted lines) and our modifications of
this model, as indicated. Observations are taken from Norberg et al. (2002) and Madgwick
et al. (2002) and are shown as crosses and without errorbars. 
The split into blue and red galaxies has been done at
$B-V=0.8$.}
\label{fig:lf}
\end{figure}

In Fig. \ref{fig:lf}, we compare the
total $b_{J}$-band luminosity function (LF hereafter, top panel) and
the red  $b_{J}$-band LF (bottom panel) for the different models to 
observations. 
Galaxies are classified as red if $B-V <$
0.8, following Croton et al. (2006). 
Observations are taken from
Norberg et al. (2002) and Madgwick et al. (2002). Again, the impact of
decreasing AGN feedback in our modifications 2) - 4) is clearly apparent.
In the bottom panel of Fig. \ref{fig:lf}, it can be seen that all models 
shown here have some problems
with reproducing the red LF. 
Interestingly,
our improved treatment of satellite stripping does not entirely solve
the overproduction of faint red galaxies noted by Croton et al. (2006).
We note that our ability to link gas stripping to the evolution of the  dark matter
mass of the subhalo will break down near
the resolution limit of the simulation. 
Right now we assume that all the gas is stripped when the subhalo is finally disrupted.
This will
likely result in effective gas stripping efficiencies that are too high for the
low mass satellites.
It thus might be advisable to use a different method for galaxies that reside in 
subhalos with  masses at infall close to the resolution limit. One possibility is to  
adopt modification 4) in this regime. Indeed, Figure 11 shows that 
modification 4) gives improved agreement with the faint end of the red LF.

Another problem becomes apparent at intermediate luminosities in our
modifications 2) - 4), where the
number of red galaxies is too low. This may be related to the
decrease in AGN feedback. Stronger dust attenuation in these galaxies
may  solve this problem, but we do not investigate this issue
in more detail here. 

Our results show that the implementation of the stripping of the satellite galaxy gas supply
in proportion to the dark matter does not impact the global properties of 
galaxies very significantly, except by decreasing the abundances of 
faint red galaxies to some degree.
However, decreasing the AGN feedback, and shutting off satellite stripping at halo masses below $10^{12}h^{-1}M_{\odot}$ in order
to better reproduce the passive fraction of central galaxies clearly
leads to inconsistencies with  the LF.
It is likely that in order
to simultaneously reproduce the passive fraction of centrals, the 
 LF of red galaxies and the bright end of the LF, a more careful retuning of the model, 
including the prescriptions for AGN feedback and dust attenuation, will
have to be carried out.

\section{Discussion}
\label{discussion}

\subsection{Treatment of satellite galaxies in the SAM}
\label{discussion_sat}

Recently, several observational and theoretical studies have shown
that the removal of the gas reservoir in satellite galaxies is
overly efficient in most
current SAMs (McCarthy  et
al.  2008; Simha  et al.  2008; Font  et al.  2008;  Weinmann et
al. 2006b; Wang et al. 2007).
In this work, we present a simple way to treat environmental effects.

It seems physically plausible that the stripping efficiency may depend
on the  mass of the host halo, on the mass of the  satellite galaxy
(e.g. Bekki 2009), on
orbital  parameters (e.g. McCarthy  et al.  2008) and on  the age  of the
Universe (e.g. Giocoli et al. 2008) 
These dependencies,
however, are not well constrained. It is not clear whether  hydrodynamical simulations 
 can properly predict the  distribution  of the (probably
multiphase) hot and ejected gas, or deal with 
the  presence of shielding magnetic fields. Smooth particle
hydrodynamics codes may have difficulties 
dealing with  dynamical instabilities  occurring  in 
regions with steep gas density gradients 
(e.g.  Agertz et
al. 2007).

In this work, we suggest using the 
approximation  that the diffuse gas halo of the satellite
(which includes the hot and ejected gas) is stripped at the same rate as the
dark matter halo. This is a very simple model which does not treat the
detailed physics of gas
stripping, but   is easy to
implement. 
We find that observations at $z$=0 can also 
be  well reproduced 
by a simple recipe in which  10-20 $\%$ of the initial gas reservoir
in the halo is stripped  per Gyr.  
We stress that in order to reproduce observations, the 
the diffuse gas must be stripped linearly, and  not exponentially. 

It is not yet clear whether 
environmental effects at higher
redshift are correctly reproduced by this method.  We find
that the dark matter stripping seems to be more efficient at high
redshift than at $z$=0. 
It will be a natural next step to compare our models to
observations of galaxy properties as a function of environment  at higher redshifts.

Eventually, it will be important to understand the nature of the gas that is found around satellite 
galaxies. ``Starvation'' is usually defined as the
stripping of the hot gas halo (Larson, Tinsley \& Caldwell 1980). However, below a mass
limit of around $10^{12} h^{-1} M_{\odot}$, 
gas cools rapidly and thus does not stay in a hot gas reservoir for a
significant amount of time. Consequently, most gas in low
mass haloes resides in the so-called ``ejected'', and not in the ``hot
phase'', with the ejected mode  composed of gas that has been expelled from the galaxy by supernovae.
The state of this ejected gas is very poorly understood:
it could consist of  cold  clumps, or it might form an extended disk, or it might
have high temperatures and low densities. The only certainty we have about this phase is that it
must remain bound to the galaxy and form a significant  part of the  reservoir for 
future star formation in low mass systems.  At low stellar masses,
stripping of this
ejected phase
is the main driver of environmental effects in the local Universe. 

In  this work,  we have  not  considered the  abundances of  satellite
galaxies as  a function  of halo mass  or the correlation  function on
small scales;  in order to  reproduce these observations, it is  likely that
tidal  disruption and  satellite-satellite  mergers have  to be  taken
into account, as SAMs tend to overproduce the abundance of satellite
galaxies (e.g. Liu et al. 2009).
  Tidal  disruption  might  also  help  with  reproducing  the
bright end of the luminosity function, as the number of mergers would
be  reduced, and would explain the presence of significant
intracluster light (Zibetti et al. 2005).
Suggestions of how tidal stripping might be implemented into SAM can be found 
in
Kim et  al. (2009), Wetzel  \& White  (2009) and
Henriques \& Thomas (2009).
Also, we have not considered the stellar ages and metallicities of 
satellites, for 
which Pasquali et al. (2009b) have found some interesting  
discrepancies between
SAMs and observations. These may help to further constrain the physics
of satellite stripping in future work.

\subsection{Comparison to a simple model of ram-pressure stripping}
\label{sec:ram}

Gunn \& Gott (1972) were the first to suggest that the ram pressure
of the hot intracluster medium may strip gas from galaxies. Ram-pressure 
stripping of the \emph{cold} disk gas has since 
received considerable theoretical
attention (e.g. Abadi et al. 1999; Schulz \& Struck 2001; Roediger
\& Br\"uggen 2007). Ram-pressure stripping of the hot gas halo, however,
has been investigated by only a few studies (e.g.
Mori \& Burkert 2000; Hester 2006; McCarthy et al. 2008).
  McCarthy et al. (2008) find that
ram-pressure stripping is more efficient than tidal stripping
in depleting the hot gas halo of most satellite galaxies.
 Our simple prescription described in section \ref{models} does not explicitely
model this effect. In this section, we test whether 
we can reproduce observations by implementing ram-pressure stripping
in a physically motivated fashion.

Our model is based on the model of ram-pressure stripping by 
Font et al. (2008), but with a few 
important modifications. We follow the orbits of the
satellites in detail, and allow stripping at every timestep, while
Font et al. (2008) only allow stripping at infall and at each time
the host halo doubles in mass.
We assume that the ram-pressure stripping of the hot gas
proceeds from the  outside  in, and does not affect the central density and hence the 
cooling efficiency of the
satellite hot gas\footnote{Unlike Font et al. (2008), we do not limit the
cooling radius ($r_{\rm cool}$, see eq. \ref{eq:cool1} and \ref{eq:cool2}) by
$R_{\rm trunc}$. In eq. \ref{eq:cool1}, we replace $m_{\rm hot}$ by $m_{\rm hot} + m_{\rm
  strip}$ for satellite galaxies.}. We
allow
the assumption of outside-in stripping 
to break down if the truncation radius of the hot halo
reaches some fraction $\beta$ of the virial radius of the dark matter halo that
hosts the satellite at the time of infall.
At this point, we allow the density profile of the hot gas to relax to an isothermal distribution
out to the virial radius $R_{\rm
  vir}$, and we strip all the ejected gas.

We describe the detailed implementation of the model in the
  L-galaxies code. While $R_{\rm trunc} > \beta R_{\rm vir}$,
we describe the density profile of the hot gas for satellites 
with a modified version of eq. \ref{eq:iso}, following Font
et al. (2008):
\begin{equation}
\label{eq:iso2}
  \rho_g(r) = \left\{ \begin{array}{ll}
      \frac{m_{\rm hot}+m_{\rm strip}}{4\pi R_{\rm vir}r^{2}} &\rm{~if~} r < R_{\rm trunc}\\
      0 & \rm{~if~} r > R_{\rm trunc}
  \end{array}\right.
\end{equation}
where  $m_{\rm strip}$ is the sum of all hot gas which has been stripped
from the galaxy since it became a satellite. $R_{\rm trunc}$ is set to $R_{\rm vir}$
when  a galaxy first changes from  central  to satellite.
(Note that  $R_{\rm vir}$ and $V_{\rm vir}$ for satellites are fixed at
infall and then remain constant.)

The stripping radius $R_{\rm strip}$ is calculated according to Gunn
\& Gott (1972) and McCarthy et al. (2008) as the radius where the ram 
pressure exceeds the pressure induced by the self-gravity of the 
satellite. For a singular isothermal sphere of hot gas, we
determine $R_{\rm strip}$ with
\begin{equation}
\alpha \rho_{\rm g, sat} (R_{\rm strip})V_{\rm vir,sat}^2 = \rho_{\rm g, cen}(R)V^2
\label{eq:ram}
\end{equation}
with $R$ and $V$ indicating the distance of the satellite to the cluster center, and its velocity
relative to it. For the pressure exerted by the self-gravity of the satellite,
we use a prefactor $\alpha = 2$, following McCarthy et al. 
(2008). 
 However, we note that removing this prefactor has very
little impact on the results.
If $R_{\rm strip} < R_{\rm trunc}$, we remove all the mass enclosed
between those two radii, and set $R_{\rm trunc} = R_{\rm strip}$.

If $R_{\rm trunc} < \beta R_{\rm vir}$, we reset
$m_{\rm strip}=0$ and $R_{\rm trunc}=R_{\rm vir}$, which means that
the hot gas 
gets diluted out to
$R_{\rm vir}$, leading to strongly decreased cooling, and increased
vulnerability to stripping. Galaxies often fluctuate between the
central and satellite status. Whenever a satellite galaxy becomes a
central again, its $R_{\rm vir}$ is updated according to its current
dark matter subhalo. $R_{\rm trunc}$ is set to $R_{\rm vir}$.
After each stripping event, the hot gas mass in the satellite
 will be modified by 
cooling, ejection, SN feedback and reincorporation. 
Before we calculate the stripping occurring in the next timestep, 
we reset
the truncation radius such that
\begin{equation}
M (r < R_{\rm trunc}) = M_{\rm hot}.
\end{equation}

\begin{figure}
\centerline{\psfig{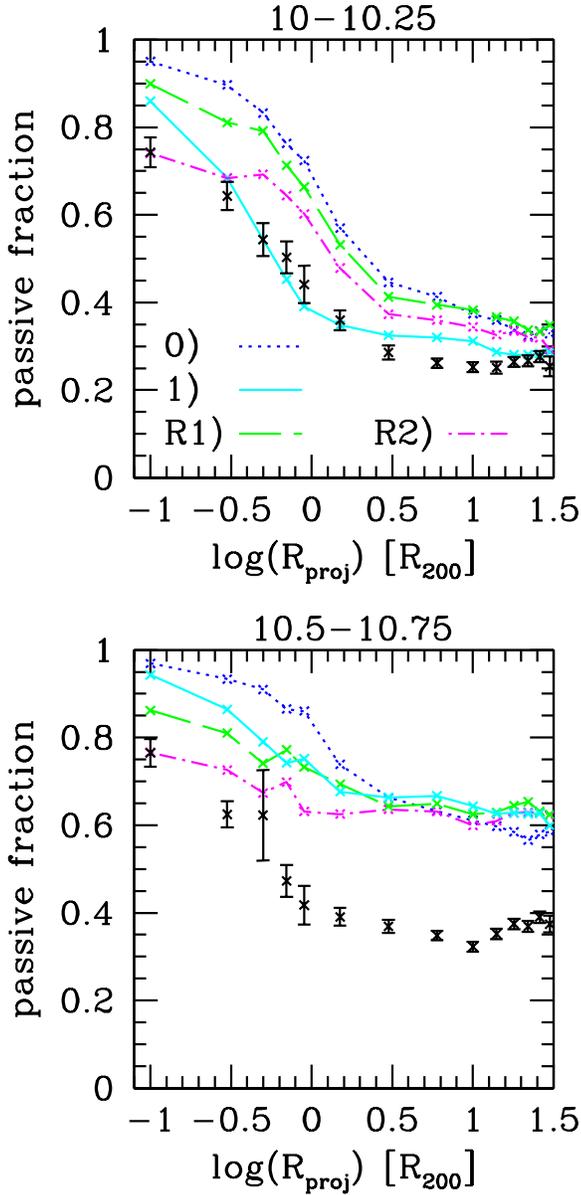}}
\caption{Passive
  fraction as a function of cluster-centric radius at stellar masses
of $\log(M/h^{-1}M_{\odot})$=10 - 10.25 (top panel) and 10.5 - 10.75 (bottom panel).
  Different linestyles and colours
  correspond to DLB07 (blue dotted lines), our model 1) (cyan solid lines)
and the two models R1) and R2) including ram-pressure stripping 
and no tidal stripping (green dashed lines and
magenta dot-dashed lines) as
  indicated and described in the text. Crosses with errorbars show the
observational results. Only clusters with masses above
$10^{14}h^{-1}M_{\odot}$
are used.}
\label{fig:rad_ram}
\end{figure}

\begin{figure}
\centerline{\psfig{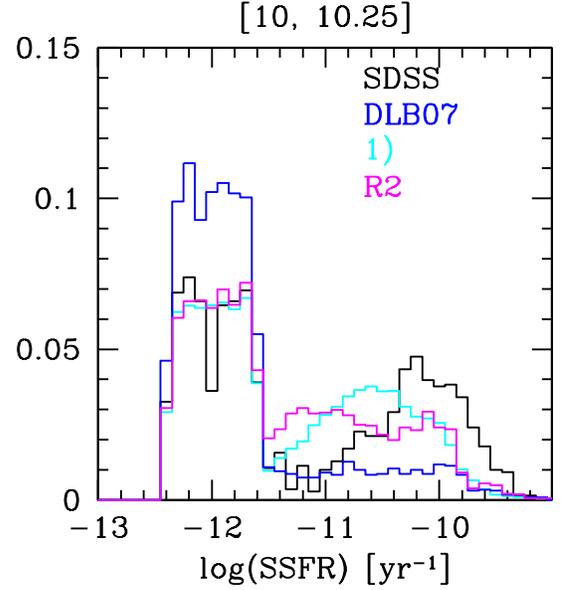}}
\caption{The distribution of satellite SSFR at stellar masses
of $\log(M/h^{-1}M_{\odot})$=10 - 10.25.
  Different colours
  correspond to DLB07 (blue lines), our model 1) (cyan lines)
and the ram-pressure stripping model R2) as
  indicated and described in the text. Black lines show the
observational results. Only clusters with masses above
$10^{14}h^{-1}M_{\odot}$
are used.}
\label{fig:ssfr_ram}
\end{figure}

Up to now, we have only discussed the ram-pressure stripping of the
hot
gas component. As shown in Fig. \ref{fig:gas}, the ejected gas mass exceeds the hot gas mass in low
mass galaxies, and it is thus 
of crucial importance how its stripping is modelled. Here, we follow 
Font et al. 
(2008) and simply assume that in a given
stripping event, the stripped
fraction of ejected gas equals
the stripped fraction of hot gas.  If $R_{\rm trunc} <
\beta R_{\rm vir}$, we assume that all the ejected gas is stripped.

We now implement the above model into a SAM based on DLB07, which
  includes the first three modifications in section 5.2. No
  tidal stripping is included, and the AGN feedback efficiency is set to be the
same as  in DLB07 for easier comparison. We compare two 
different models: R1) and R2). In model R1), we use $\beta$= 0.1, and in 
model R2), $\beta=$0.01.
Results for the stellar mass bins $\log(M/h^{-1}M_{\odot})$ = 10-10.25 (top
panel) and  $\log(M/h^{-1}M_{\odot})$ = 10.5-10.75 (bottom panel)
are shown as green dashed lines for model R1) and as magenta
dot-dashed 
lines for model R2) in Fig \ref{fig:rad_ram}. For comparison, we
also show results 
from DLB07 and our model 1), where the diffuse gas is stripped at the same rate
as the dark matter.

For low mass galaxies, we find that the ram-pressure is much more efficient
at stripping out the gas than model 1), and both models 
R1) and R2) produce passive fractions that are too high to match
observations.\footnote{We have also tested a model with $\beta$=0, which means that the
central density of the hot gas is never affected by stripping. In this
case, satellite galaxies have as high star formation rates as central
galaxies. Interestingly, this indicates that it is not the reduced hot
gas mass, but the reduced central density of the hot gas, which causes
environmental effects in all our models.}
On the other hand, in high mass galaxies, the ram-pressure effects are 
weaker than the tidal effects, and the ram-pressure models cannot produce a strong enough
change in passive fraction as a function of clustercentric radius.
 In Fig. \ref{fig:ssfr_ram}, we show the distribution 
of SSFR for the stellar mass bin $\log(M/h^{-1}M_{\odot})$ = 10-10.25. In magenta, we show results for
the model R2). Clearly, there are too many
galaxies
with intermediate SSFR in this model.

In summary, we  find three problems with a model in which environmental effects 
arise solely due to ram-pressure stripping. First, stripping  effects in 
clusters for low mass galaxies are too strong to explain the observations. Second, stripping effects 
are not strong enough to reproduce the observations for high mass galaxies. Third, there
are too many low mass  galaxies with intermediate SSFR.
The first problem could indicate that our understanding of the so-called
``ejected phase'', which is important for low mass galaxies, is inaccurate.
Although the ejected gas is currently unavailable for star formation, 
it maybe  nevertheless strongly bound to the galaxy, and difficult to strip.
Another potential reason for the problem 
  is that SAMs like DLB07 are likely
to 
overestimate the hot gas mass in groups with masses 
between $10^{12}$ and $10^{14} h^{-1}M_{\odot}$,
 especially in the central, most
dense regions of these systems (Bower et al. 2008). This
could lead to too strong pre-processing of cluster galaxies in groups.
The second problem could indicate that tidal effects
are in fact important for higher mass galaxies.
By decreasing the dark matter content in subhaloes, 
tidal effects make
high mass galaxies more vulnerable to ram-pressure stripping, but also to
AGN feedback. 
The third problem indicates that ram-pressure stripping as implemented here 
does not
produce a bimodal satellite galaxy population. 
A similar problem seems
to be present in the ram-pressure stripping model of Font et
al. (2008), as pointed out by Balogh et al. (2009). This could also 
hint at the importance of tidal effects, which we have shown 
to produce bimodal distribution in the SSFR of galaxies.

We note that Font et
al. (2008) have found that their ram-pressure model can reproduce the
colours of satellite galaxies well. As they do not follow the orbits
of
individual satellites in detail, and use different assumptions for the
hot gas profiles and the treatment of cooling in satellites, it is hard to directly compare the
two models and to pinpoint the reason for this discrepancy.

\section{Summary}
\label{summary}

We explore a set of modifications to the SAM of  
DLB07 with the goal to reproduce (i) the passive fractions as a
function of cluster-centric radius out to the field, (ii) the
detailed distribution of the SSFRs of satellite galaxies and
(iii) the passive fractions of central galaxies as a function of 
stellar mass. We apply a
realistic cluster finder to the SAM to allow a fair
comparison with observational results from the cluster catalogue
of vdL07.
Our main results are the following.

\begin{itemize} 

\item A model in which the diffuse gas is stripped at the same rate
that the dark matter subhalos lose mass due to tidal stripping can  
reproduce observations reasonably well.
 ``Starvation'' is usually defined as the fast removal of the hot gas halo
surrounding satellite galaxies (Larson, Tinsley \& Caldwell 1980) due
to both tidal and ram-pressure stripping. In this work, we find
indications that this picture is in need of revision. 
First, stripping of the diffuse gas reservoir 
is slow. Second, it is crucial for the success of our model 
that even the gas ejected by supernovae is
not stripped instantaneously, but viewed as part of the
overall gas reservoir.

\item After infall, the median  subhalo identified at $z$=0 loses 15--20$\%$  of
  its initial dark
  matter halo mass per Gyr, which implies that tidal stripping
  is typically only complete after 5--7 Gyrs. 
 Dark matter stripping seems to proceed linearly, and not
  exponentially. This is  crucial for reproducing the observed
  bimodality in the specific star formation rates of satellites
  in our model. We also find that 
stripping of dark matter
  becomes more efficient at high redshift.

\item We present a simple method for implementing gas stripping in
  SAMs which do not follow the dark matter subhaloes
  of satellites. Our prescription is that    10--20\% of the hot and
  ejected gas recorded at infall should be stripped  per Gyr. This model gives good agreement
  with observations at $z$=0.

\item A model which only includes a physically motivated prescription
for ram-pressure stripping of the hot gas in satellite galaxies seems
to lead to an overproduction of low mass passive galaxies in clusters, but
also too little environmental effects for high mass galaxies. 

\item We have found that DLB07 do not reproduce the passive fraction
  of central galaxies as a function of stellar mass in detail. 
In particular,
DLB07 predict too few passive central galaxies with low masses, and
too many with high masses. We find that we can improve 
agreement with observations by reducing the AGN feedback efficiency,
and by assuming that  satellite galaxies  in host haloes with
masses below $10^{12}
h^{-1}M_{\odot}$ retain  all their hot and ejected gas.
However, the reduction of AGN feedback leads to an overproduction 
of the total number of massive and luminous galaxies.
In order to simultaneously reproduce the low abundance, and the relatively 
high active fraction of massive galaxies, the form
of the AGN feedback in DLB07 may need to be changed.
One possible way to solve this problem might be to introduce some stochasticity into the
shut-down
of cooling at high halo masses, for example by taking into account the
effect of major mergers or morphological quenching (Hopkins et
al. 2008; Martig et al. 2009).
We also find that reproducing both the red luminosity function 
and the passive fraction of central and satellite galaxies
is surprisingly difficult, 
indicating either an inconsistency between different observations, or
a potential problem with the dust treatment in the SAM.

\end{itemize}

\section*{Acknowledgments}
SW thanks Eyal Neistein for a careful reading of the manuscript and
many helpful comments; Qi Guo, Andreea Font, Simon White, 
Richard Bower, Barbara Catinella, Andreas Faltenbacher,
Ramin Skibba, Yan-Mei Chen, Andrea Macci\`{o} and Xi
Kang for
useful discussion, and Volker Springel and Mike Boylan-Kolchin for their
help with the Millennium database. We also thank the anonymous referee
for the suggestion to add a model for ram-pressure stripping.
GDL acknowledges financial support from the European Research Council
under the European Community's Seventh Framework Programme (FP7/2007-2013)/ERC
grant agreement n. 202781.

\end{document}